\documentclass[fleqn,usenatbib]{mnras}

\usepackage{amssymb}	

\usepackage{newtxtext,newtxmath}

\usepackage[T1]{fontenc}
\usepackage{lipsum}

\DeclareRobustCommand{\VAN}[3]{#2}
\let\VANthebibliography\thebibliography
\def\thebibliography{\DeclareRobustCommand{\VAN}[3]{##3}\VANthebibliography}

\usepackage{graphicx}	
\usepackage{amsmath}	
\usepackage{hyperref}   
\usepackage{natbib}

\usepackage{scalerel}
\usepackage{tikz}
\usetikzlibrary{svg.path}
\definecolor{orcidlogocol}{HTML}{A6CE39}
\tikzset{
  orcidlogo/.pic={
    \fill[orcidlogocol] svg{M256,128c0,70.7-57.3,128-128,128C57.3,256,0,198.7,0,128C0,57.3,57.3,0,128,0C198.7,0,256,57.3,256,128z};
    \fill[white] svg{M86.3,186.2H70.9V79.1h15.4v48.4V186.2z}
                 svg{M108.9,79.1h41.6c39.6,0,57,28.3,57,53.6c0,27.5-21.5,53.6-56.8,53.6h-41.8V79.1z M124.3,172.4h24.5c34.9,0,42.9-26.5,42.9-39.7c0-21.5-13.7-39.7-43.7-39.7h-23.7V172.4z}
                 svg{M88.7,56.8c0,5.5-4.5,10.1-10.1,10.1c-5.6,0-10.1-4.6-10.1-10.1c0-5.6,4.5-10.1,10.1-10.1C84.2,46.7,88.7,51.3,88.7,56.8z};
  }
}
\newcommand\orcidicon[1]{\href{https://orcid.org/#1}{\mbox{\scalerel*{
\begin{tikzpicture}[yscale=-1,transform shape]
\pic{orcidlogo};
\end{tikzpicture}
}{|}}}}

\newcommand{\eagle}{{\sc{eagle}}}
\newcommand{\RefAGN}{RefL50}
\newcommand{\RefSN}{RefL25}
\newcommand{\Recal}{RecalL25}
\newcommand{\WeakFB}{WeakFB}
\newcommand{\StrongFB}{StrongFB}
\newcommand{\NoAGN}{NoAGN}
\newcommand{\AGNdT}{AGNdT9}
\newcommand{\NoAGNfp}{NoAGN-fp}
\newcommand{\NoAGNfptg}{NoAGN-fp-tot-gas}

\defcitealias{Lara-Lopez2019}{LL19}

\title[Effective yields as tracers of feedback]{Effective yields as tracers of feedback effects on metallicity scaling relations in the EAGLE cosmological simulations}

\author[M. C. Zerbo et al.]{
M.C.~Zerbo,$^{1,2,3,\orcidicon{0009-0000-5717-267X}}$\thanks{E-mail: candelazerbo@gmail.com}
M.E.~De Rossi,$^{2,3,\orcidicon{0000-0002-4575-6886}}$\thanks{E-mail: mariaemilia.dr@gmail.com}
M.A.~Lara-López,$^{4,5,\orcidicon{0000-0001-7327-3489}}$
S.A.~Cora,$^{1,6,\orcidicon{0000-0002-0841-9127}}$
and L.J.~Zenocratti$^{1,6,\orcidicon{0000-0001-8271-1794}}$
\\
$^{1}$Facultad de Ciencias Astronómicas y Geofísicas, Universidad Nacional de La Plata, Paseo del Bosque s/n, B1900FWA, La Plata, Argentina\\
$^{2}$Universidad de Buenos Aires, Facultad de Ciencias Exactas y Naturales, Buenos Aires, Argentina\\
$^{3}$CONICET-Universidad de Buenos Aires, Instituto de Astronomía y Física del Espacio (IAFE), Buenos Aires, Argentina\\
$^{4}$Departamento de Física de la Tierra y Astrofísica, Universidad Complutense de Madrid, E-28040 Madrid, Spain\\
$^{5}$Instituto de Física de Partículas y del Cosmos IPARCOS, Fac. de Ciencias Físicas, Universidad Complutense de Madrid, E-28040, Madrid, Spain\\
$^{6}$Instituto de Astrofísica de La Plata, CONICET-Universidad Nacional de La Plata, Paseo del Bosque s/n, B1900FWA, La Plata, Argentina
}

\date{Accepted 2024 February 12. Received 2024 February 12; in original form 2023 September 13}

\pubyear{2023}

\begin{document}
\label{firstpage}
\pagerange{\pageref{firstpage}--\pageref{lastpage}}
\maketitle

\begin{abstract}
Effective yields, $y_{\rm eff}$, are defined by fundamental galaxy properties (i.e., stellar mass -$M_{\star}$-, gas mass -$M_{\rm gas}$- and gas-phase metallicity). For a closed-box model, $y_{\rm eff}$ is constant and equivalent to the mass in metals returned to the gas per unit mass locked in long-lived stars. Deviations from such behaviour have been often considered observational signatures of past feedback events. By analysing \eagle \, simulations with different feedback models, we evaluate the impact of supernovae (SN) and active galactic nuclei (AGN) feedback on $y_{\rm eff}$ at redshift $z=0$. When removing supermassive black holes (BH) and, hence, AGN effects, in simulations, galaxies are located around a plane in the $M_{\star} - M_{\rm gas} - {\rm O/H}$ parameter space (being O/H a proxy for gas metallicity, as usual), with such a plane roughly describing a surface of constant $y_{\rm eff}$. As the ratio between BH mass and $M_{\star}$ increases, galaxies deviate from that plane towards lower $y_{\rm eff}$ as a consequence of AGN feedback. For galaxies not strongly affected by AGN feedback, a stronger SN feedback efficiency generates deviations towards lower $y_{\rm eff}$, while galaxies move towards the opposite side of the plane (i.e., towards higher values of $y_{\rm eff}$) as SN feedback becomes weaker. Star-forming galaxies observed in the Local Universe are located around a similar 3D plane. Our results suggest that the features of the scatter around the observed plane are related to the different feedback histories of galaxies, which might be traced by $y_{\rm eff}$.
\end{abstract}

\begin{keywords}
galaxies: abundances -- galaxies: evolution -- galaxies: formation -- galaxies: fundamental parameters -- galaxies: star formation -- methods: numerical
\end{keywords}



\section{Introduction}
\label{sec:intro}

The simplest model proposed for the metal enrichment of a galaxy is the so-called `closed-box', which assumes that the system is isolated from its environment. Clearly, there is no mass flow towards or outwards the system and, as a consequence, the baryonic mass remains constant with time. Following the prescriptions of a closed-box model, the evolution of a system comes down to the synergy between the stellar and gas components, whose variations are regulated only by star formation and stellar yields (\citealt{Pagel1975}). On the one hand, stars are created after the collapse of molecular clouds at a certain rate that depends on the availability of cold gas reservoirs. On the other hand, as a fair number of stars in each generation reaches the end of their lifetime, successive events of supernovae (SN) and stellar winds enrich the interstellar medium (ISM) with heavier elements. If we further assume that the ejected material from stars is homogeneously and instantaneously mixed, the evolution of the metallicity of the gas obeys a simple analytical expression:
\begin{equation}
\label{eq:Z_evolution}
\Tilde{Z}_{\rm gas}(t) =  - y_Z \ln \left[ \Tilde{\mu}(t) \right], 
\end{equation}
where $y_{\rm Z}$ denotes the true stellar yield, defined as the mass of newly produced metals (via nucleosynthesis) expelled by a generation of stars with regard to the total mass that remains inside of long-lived stars and compact remnants. $\Tilde{Z}_{\rm gas}(t)$ and $\tilde{\mu} (t)= \tilde{M}_{\rm gas}(t)/(\tilde{M}_\star(t) + \tilde{M}_{\rm gas}(t))$ are the gas metallicity and gas mass fraction at time $t$, respectively, where $\Tilde{M}_\star$ is the stellar mass.

There is abundant observational evidence that a closed-box behaviour does not constitute a suitable formation scenario for most galaxies. Instead, a more appropriate description can be achieved considering the interaction between them and the surrounding environment, through inflows and outflows of mass (e.g. \citealt{Edmunds1984}; \citealt{Edmunds1990}; \citealt{Tremonti2004}; \citealt{Dalcanton2007}; \citealt{Erb2008}; \citealt{Tortora2022}). In this context, by comparing the predictions of a closed-box model with the behaviour of real galaxies, different works have tried to address the relative impact of feedback effects (e.g. outflows) on galaxies of different masses (e.g. \citealt{Lara-Lopez2019}).  Nevertheless, such approach relies on the validity of other closed-box model approximations. Regarding the instantaneous mixing assumption, it is expected to be accurate when the metallicity of galaxies is estimated by means of the oxygen abundance in HII regions \citep[e.g.][]{Dalcanton2007}. Since the production of oxygen is predominantly related to winds of massive stars and SN events, a short time scale for recycling mixing is a reasonable approximation. As O is the most common gas-phase metallicity tracer, such assumption is generally valid. With respect to perfect mixing, the existence of metallicity gradients within galaxies indicate that mixing scales seem to be long.  Nevertheless, it is expected that the latter issues lead to smaller deviations from the closed-box model than those generated by gas flows \citep[e.g.][]{Tremonti2004}. We notice, however, that some works suggest that gas-rich dwarfs might require a more careful inspection (e.g. \citealt{Werk2011}).

If we invert equation~(\ref{eq:Z_evolution}) and evaluate the expression using the gas metallicity ($Z_{\rm gas}$) and gas mass fraction ($\mu = M_{\rm gas}  / (M_\star + M_{\rm gas}$), being $M_\star$ the stellar mass) of observed galaxies, we can define the effective yield  as usual (e.g., \citealt{Dalcanton2007}):
\begin{equation}
\label{eq:yeff}
y_{\rm eff} = \frac{Z_{\rm gas}}{\ln(1/\mu)}.
\end{equation}
The effective yield is constant and equal to the true stellar $y_{\rm Z}$ for galaxies that evolve as closed-box systems. On the contrary, \cite{Edmunds1990} demonstrated that inflows and outflows of material lead to $y_{\rm eff} \leq y_{\rm Z}$ as a result of metal-enriched outflows and/or the accretion of primordial gas. The only exception that can generate the opposite trend is the accretion of gas with metallicity similar to or higher than the system. Nevertheless, this situation is highly unlikely and it is usually not considered. Furthermore, \cite{Dalcanton2007} showed that in gas-rich galaxies the most effective mechanism for lowering the value of effective yields are metal-rich outflows. Conversely, metal-poor accreted gas does not have a significant impact on the effective yield of galaxies with high gas mass fraction (\citealt{Dalcanton2007}). In this case, even though the accretion will lower the gas metallicity, $Z_{\rm gas}$, the gas mass fraction will increase as well, leaving $y_{\rm eff}$ almost unaffected. This particular scenario is known as a pseudo-closed-box equilibrium. 

Hydrodynamical simulations have proved to be powerful tools for explaining observations and gaining more insight into the intertwined processes that occur during the evolution of real galaxies (e.g. \citealt{Ma2016}; \citealt{Torrey2019}). In particular, it has been shown that feedback processes powered up by SN events and AGN are essential in shaping galaxy metallicity scaling relations \cite [e.g.][]{DeRossi2017}. In general, different works based on simulations show that SN feedback plays an important role on the regulation of the star formation activity of low-mass galaxies, generating a decrease in their star formation rate (SFR) and, hence, in their metallicity \citep[e.g.][]{Brooks2007}.

State-of-the art cosmological simulations predict also a critical role of AGN feedback on the determination of metallicity scaling relations of massive galaxies. For example, \citet{DeRossi2017} carried out a detailed analysis of the ‘Evolution and Assembly of GaLaxies and their Environments’ (\eagle, \citealt{Schaye2015}; \citealt{Crain2015}) suite of cosmological hydrodynamical simulations,  showing that they are able to broadly describe the observed flattening of the mass-metallicity relation (M$_\star$Z$_{\rm g}$R) for massive galaxies in the Local Universe.\footnote{We note, however, that the exact normalisation and shape of the M$_\star$Z$_{\rm g}$R are still debated. Different observational methods for inferring the key galaxy properties yield significantly different answers \citep[][]{Telford2016} and, hence, the comparison between simulations and observations is not straightforward.} These authors reported that AGN feedback plays a central role in regulating the chemical evolution of massive galaxies, driving such behaviour. The primary cause of these effects appears to be the energy and momentum released by AGN, leading to a depletion of cold gas reservoirs in their host galaxies by heating and/or the ejection of metal-enriched material. Consequently, the process inhibits both star formation and chemical evolution.

In the last decades, a key relationship between $y_{\rm eff}$ and the baryonic mass, $M_{\rm bar} = M_\star + M_{\rm gas}$, has been widely studied. At the low-mass end, several works have reported that effective yields increase with baryonic mass (\citealt{Garnett2002}; \citealt{Tremonti2004}; \citealt{Lee2006}; \citealt{Ekta2010}). Three different plausible channels have been often proposed to explain the decrease of $y_{\rm eff}$ towards lower $M_{\rm bar}$: the more efficient removal of metals, via galactic winds, from shallower potential wells (\citealt{Garnett2002}; \citealt{Tremonti2004}; \citealt{Silich2001}); the infall of pristine gas (\citealt{SanchezAlmeida2014,SanchezAlmeida2015}); and, a higher ISM mixing efficiency, driven by the migration of metal-poor gas from the galaxy outskirts towards the more metal-enriched central region (\citealt{Ekta2010}). As higher masses are considered, \cite{Tremonti2004} first reported a change in the behaviour of observed galaxies at $M_{\rm bar} \gtrsim 10^{10}~\rm{M}_\odot$, which show a flatter $M_{\rm bar} - y_{\rm eff}$ relation, on average. According to \cite{DeRossi2017}, \eagle~simulations predict that $y_{\rm eff}$ tends to decrease above a similar characteristic mass. In the simulations, such a trend appears to be the result of the cumulative effects of previous AGN feedback, which operate through the following mechanisms: heating the star-forming gas component, thereby suppressing star formation and leading to a passive galaxy, as well as expelling metal-enriched material through galactic outflows.

\citetalias{Lara-Lopez2019} performed a detailed comparison between the observed $M_{\rm bar} - y_{\rm eff}$ relation and that obtained from \eagle~at $M_\star > 10^9~\rm{M}_\odot$, which is consistent with the mass range studied by \cite{DeRossi2017}. \citetalias{Lara-Lopez2019} reported, for the first time, an anti-correlation at the high-mass end for both observed and simulated galaxies. In addition, their results showed a clear bimodal behaviour when galaxies are separated by stellar age. On the one hand, 
younger galaxies present higher values of $y_{\rm eff}$ as we consider higher masses. They also exhibit higher gas mass fractions, specific star formation rates sSFR and lower star formation efficiencies (SFE = SFR/$M_{\rm gas}$). On the other hand, old galaxies tend to have high $M_{\rm bar}$ and, as the considered stellar age increases, the $M_{\rm bar} - y_{\rm eff}$ relation becomes flatter until an anti-correlation appears. This population is characterised by low gas fractions, sSFR and SFE, that is to say, is composed of passive galaxies whose star formation has been quenched.

By using \eagle~simulations, in this article, we provide new insights about the origin of the $M_{\rm bar} - y_{\rm eff}$ relation, considering different feedback models.  We also try to assess the capability of $y_{\rm eff}$ to diagnose the accumulated effects of SN and/or AGN feedback processes on 2D and 3D metallicity scaling relations. Given that the \eagle~ set of simulations are publicly available and offer results from different models of SN and AGN feedback efficiencies, they are suitable for our study. Furthermore, \eagle~ simulations were used in the work upon which we based our study, \cite{Lara-Lopez2019} (hereafter, \citetalias{Lara-Lopez2019}), and show very good agreement with the observed values of $y_{\rm eff}$. Our paper is divided into the following sections. In Section~\ref{sec:simulations}, we present a brief description of \eagle~ simulations and our galaxy sample. In Section~\ref{sec:2d_relations}, we explore the role of SN and AGN feedback processes on $y_{\rm eff}$ and other related quantities, such as oxygen abundance, stellar mass and gas fraction. In Section~\ref{sec:3d_relations}, we analyse the 3D scaling relation defined by $M_\star$, oxygen abundance and gas mass, exploring the impact of feedback on its features. We discuss results from simulations and perform a comparison with observations in Section~\ref{sec:discussion}. Finally, our conclusions are summarised in Section~\ref{sec:summary}.

\section{The {\sc{EAGLE}} simulations}
\label{sec:simulations}
The \eagle \, cosmological hydrodynamical simulations were run with a modified version of the {\sc{treepm-sph}} {\sc{gadget3}} code (\citealt{Springel2005}). The joint evolution of dark matter and baryons are tracked within cosmological representative volumes, considering different periodic co-moving boxes, mass resolutions, and sub-grid physics models. The sub-grid prescriptions take into account unresolved processes, such as star formation, stellar evolution, radiative cooling, photoionization heating, metal enrichment and feedback associated with massive stars. Most \eagle \, models include seeding, growth and merging of supermassive black holes (BH), and AGN feedback; see \citet{Schaye2015} and \citet{Crain2015}, for full details.  

A $\Lambda$CDM cosmology is adopted, with parameters consistent with \citet{Planck2015}, namely $\Omega_{\Lambda} = 0.693$, $\Omega_{\rm m} = 0.307$, $\Omega_{\rm b} = 0.0483$, $n_{\rm s} = 0.961$, $Y = 0.248$, and $h = 0.677$, where symbols have their usual meaning. A glass-like particle initial configuration was implemented as initial condition, with a second-order Lagrangian perturbation following \citet{Jenkins2010}, using the public Panphasia Gaussian white noise field (\citealt{Jenkins2013}). Full details about the generation of the initial conditions can be found in appendix B of \citet{Schaye2015}.

Dark matter haloes are identified using the Friends-of-Friends algorithm (FoF, \citealt{Davis1985}), and baryonic particles are assigned to the same FoF halo as their nearest dark matter neighbour. Galaxies containing baryons and dark matter are then identified with the {\sc{subfind}} algorithm (\citealt{Springel2005}; \citealt{Dolag2009}). The galaxy that hosts the most bound dark matter particle in a halo is defined as the central galaxy of that halo, and the remaining subhalos are classified as satellite galaxies.\footnote{Properties of \eagle \, galaxies and haloes can be queried through the public \eagle \, database \citep{McAlpine2016}. The \eagle \, particle data is  also available \citep{eagle2017}.}

Within the \eagle \, suite, different simulations are identified according to the linear co-moving extent $L$ of the simulated cubic volume and the particle count $N$. For example, a simulation with label L0025N0376 corresponds to a cubic box with 25 co-moving megaparsecs (cMpc) of side length, performed with $376^3$ dark matter particles and an equal initial number of gas particles. Also, the complete label of a given \eagle \, simulation includes a prefix that indicates the sub-grid model adopted. The reference model (`Ref' prefix) was calibrated to reproduce $z\approx0$ observations (see Section~\ref{sec:physics}). A recalibrated model (`Recal' prefix) was also considered, which improves the agreement with observational data for the highest resolution simulations within \eagle \, suite. In this article, we particularly analyse the simulations so-called RefL0025N0376 and RefL0050N0752, in order to compare them with runs corresponding to different feedback parameters but similar $L - N$ combinations. For the sake of simplicity, the former simulations will be regarded as `\RefSN' and `\RefAGN', respectively. Variations with respect to the reference model were tested by modifying one or more parameters of the sub-grid modules (\citealt{Crain2015}). In this work, in addition to the reference case, we analyse models that assume: a weaker and a stronger efficiency of SN feedback (`\WeakFB' and `\StrongFB' models, respectively), no AGN feedback (`\NoAGN' model), and an enhancement of the gas temperature increase due to AGN feedback ($\Delta T_{\rm{AGN}} = 10^9~\rm{K}$; i.e. `\AGNdT' model); see Section~\ref{sec:physics}, for details. Simulations that apply variations of SN and AGN feedback parameters were run with a similar numerical resolution, but using simulated boxes of $L = 25~{\rm cMpc}$ and $L = 50~{\rm cMpc}$, respectively.  Hence, given the smaller volume, the former simulations include a lower number of galaxies. In the following sections, we summarise the main sub-grid models and physical parameters included in \eagle \, that are relevant for our study. For a more complete explanation, the reader is referred to \citet{Schaye2015} and \citet{Crain2015}.

\subsection{Summary of {\sc{EAGLE}} subgrid implementation}
\label{sec:physics}
In this section, we briefly describe the most relevant aspects of sub-grid physics in \eagle \, simulations. In particular, we provide a detailed description of the key subgrid parameters involved in the different feedback models examined in this study, which are summarised in  Table~\ref{tab:simus}.  Parameters corresponding to the reference model were calibrated considering the following $z\approx 0$ observables: the galaxy stellar mass function (GSMF), galaxy sizes, and the relation between BH mass and stellar mass. Other models used in this work test the predictions of single-parameter variations with respect to the reference implementation.

\begin{table*}
	\centering
	\caption{Main physical parameters of the \eagle \, simulations used in this work. From left to right, the columns show the simulation identifier, side length of the co-moving volume ($L$), number of dark matter particles ($N$, which is similar to the initial number of gas particles), initial baryonic particle mass ($m_{\rm g}$), dark matter particle mass ($m_{\rm DM}$), asymptotic maximum ($f_{\rm{th,max}}$) and minimum ($f_{\rm{th,min}}$) values of $f_{\rm th}$ (see equation~\ref{eq:fth}), the parameters that control the characteristic density and the power-law slope of the density dependence of the energy feedback from star formation ($n_{\rm{H,0}}$ and $n_{\rm n}$, respectively), the (sub-grid) accretion disc viscosity parameter ($C_{\rm visc}$), the temperature increment of stochastic AGN heating ($\Delta T_{\rm AGN}$), and the number of galaxies extracted from the simulation ($N_{\rm gal}$, see Section~\ref{sec:selection}). The upper section corresponds to simulations run with the reference model, which has been calibrated to reproduce $z \approx 0$ observations (\RefSN \, and \RefAGN \, simulations). The lower section comprises simulations run with models featuring single-parameter variations with respect to the reference one. Numbers in bold indicate such variations.}
	\label{tab:simus}
	\begin{tabular}{rccccccccccr}
		\hline
		Identifier & $L$ & $N$ & $m_{\rm g}$ & $m_{\rm DM}$ & $f_{\rm th,max}$ & $f_{\rm th,min}$ & $n_{\rm H,0}$ & $n_{\rm n}$ & $C_{\rm visc}/2\pi$ & $\Delta T_{\rm AGN}$ & $N_{\rm gal}$ \\
		& [cMpc] & & [$\rm{M}_\odot$] & [$\rm{M}_\odot$] & & & [cm$^{-3}$] & & & $\log_{10}$ [K] & \\
		\hline
		{\it Calibrated models}\\
		\RefSN & 25 & $376^3$ & $1.81 \times 10^6$ & $9.70 \times 10^6$ & 3.0 & 0.3 & 0.67 & 2 / ln 10 & 10$^0$ & 8.5 & 194\\
		\RefAGN & 50 & $752^3$ & $1.81 \times 10^6$ & $9.70 \times 10^6$ & 3.0 & 0.3 & 0.67 & 2 / ln 10 & 10$^0$ & 8.5 & 1444 \\
		{\it Reference model variations}\\
		\WeakFB & 25 & $376^3$ & $1.81 \times 10^6$ & $9.70 \times 10^6$ & {\bf 1.5} & {\bf 0.15} & 0.67 & 2 / ln 10 & 10$^0$ & 8.5 & 231 \\
		\StrongFB & 25 & $376^3$ & $1.81 \times 10^6$ & $9.70 \times 10^6$ & {\bf 6.0} & {\bf 0.6} & 0.67 & 2 / ln 10 & 10$^0$ & 8.5 & 115\\
		\NoAGN & 50 & $752^3$ & $1.81 \times 10^6$ & $9.70 \times 10^6$ & 3.0 & 0.3 & 0.67 & 2 / ln 10 & {\bf ---} & {\bf---} & 1505 \\
		\AGNdT & 50 & $752^3$ & $1.81 \times 10^6$ & $9.70 \times 10^6$ & 3.0 & 0.3 & 0.67 & 2 / ln 10 & 10$^0$ & {\bf 9.0} & 1382 \\		
		\hline
	\end{tabular}
\end{table*}

\subsubsection{Star formation and chemical enrichment}
Radiative cooling, photoheating and chemical enrichment are implemented element by element following \citet{Wiersma2009a,Wiersma2009b}, tracking individually 11 chemical elements (H, He, C, N, O, Ne, Mg, Si, S, Ca, and Fe). The model also assumes an optically thin gas component in ionisation equilibrium, which is exposed to both, an ionizing UV/X-ray background (\citealt{Haardt2001}) and the cosmic microwave background. The simulations track stellar mass-losses of the aforementioned elements considering three channels: stellar winds and type II supernovae (SNII) from $M_\star > 6~\rm{M}_\odot$ stars, type Ia supernovae (SNIa) originated in catastrophic mass transfer between close binary stars, and winds from stars belonging to the asymptotic giant branch (AGB). Stellar yields that depend on the initial metal abundance are adopted: yields from \citet{Portinari1998} that consider mass-loss from massive stars were used for SNII, while yields of \citet{Marigo2001} and \citet{Wiersma2009b} were implemented for AGB stars and SNIa, respectively.

Following \citet{Schaye2008}, star formation in \eagle \, is implemented stochastically, assuming a volume density threshold of total hydrogen $n_{\rm{H}}^\ast$ that depends on metallicity $Z$ (\citealt{Schaye2015}; \citealt{Crain2015}). The hydrogen number density, $n_{\rm H}$, is related to the overall gas density, ${\rho}_{\rm g}$, considering $n_{\rm H} \equiv X {\rho}_{\rm g} / m_{\rm H}$, where $X$ is the hydrogen mass fraction and $m_{\rm H}$ is the mass of a hydrogen atom \citep{Crain2015}. A temperature floor $T_{\rm eos}$, associated with the equation of state $P_{\rm eos} \propto \rho_{\rm g}^{4/3}$, is also applied, which is normalised to $T_{\rm eos}=8\times 10^3~\rm{K}$ at $n_{\rm H}=10^{-1}~\rm{cm}^{-3}$. When gas particles fulfil the conditions $n_{\rm H} > n_{\rm{H}}^\ast$ and $\log_{10}(T/{\rm{K}}) < \log_{10}(T_{\rm eos}/{\rm{K}}) + 0.5$, they are considered star-forming (SF) gas particles, and are assigned a star formation rate (SFR) that follows the Kennicutt-Schmidt relation (\citealt{Kennicutt1998}).

\subsubsection{Energy feedback from star formation}
Thermal feedback from star formation is applied stochastically, adopting the feedback model of \citet{DallaVecchia2012}. This model carries out a stochastic selection of neighbouring gas particles that are heated by a temperature increment of $10^{7.5}~{\rm K}$. In particular, a fraction $f_{\rm th}$ of energy from core-collapse supernovae (SNII) is injected, taking into account the local metallicity and gas density. Such energy is released into the ISM $30~{\rm Myr}$ after the birth of a stellar population (\citealt{Schaye2015}; \citealt{Crain2015}), and is given by:
\begin{equation}
\label{eq:fth}
f_{\rm th}=f_{\rm{th,min}}+\frac{f_{\rm{th,max}}-f_{\rm{th,min}}}{1+\left(\frac{Z}{0.1Z_\odot}\right)^{n_Z} \left(\frac{n_{\rm H,birth}}{n_{\rm H,0}}\right)^{-n_{\rm n}}},
\end{equation}
\noindent
where $Z_\odot=0.0127$ is the solar metallicity, $n_{\rm H,birth}$ is the density inherited by the star particle from its parent gas particle, $Z$ is the metallicity, $f_{\rm th,min}$ and $f_{\rm th,max}$ are the asymptotic values of $f_{\rm th}$, and $n_Z$, $n_{\rm n}$ and $n_{\rm H,0}$ are free parameters that were chosen to reproduce the $z\approx 0$ GSMF and the galaxy mass-size relation (see \citealt{Schaye2015}, for details). In the model, $n_Z=n_{\rm n}$ is assumed. 

In \eagle, galaxy formation and evolution is governed primarily by the supply of gas into the ISM, being this process regulated by feedback. Changing the efficiency of star formation feedback has a significant impact on many galaxy properties (e.g. \citealt{Crain2015} and references therein; \citealt{DeRossi2017}). The reference \eagle \, model adopts $f_{\rm th,max}=3$ and $f_{\rm th,min}=0.3$, while, in the \WeakFB \, and \StrongFB \, models, these values are scaled by a factor of 0.5 and 2, respectively (see Table~\ref{tab:simus}). We note that the \StrongFB \, model predicts significant changes in the stellar mass fraction ($M_\star/M_{\rm h}$) as a function of halo mass ($M_{\rm h}$) when compared with the reference model  (\citealt{Crain2015}), which is the one calibrated against observations. In particular, the \StrongFB \, model predicts the formation of galaxies with significantly lower $M_\star$ and metallicities, at a given $M_{\rm h}$, than the reference prescription. In this article, the \StrongFB \, and \WeakFB \, models are considered to test the effects of varying SN feedback efficiencies on metallicity scaling relations but, given that they are not adjusted to reproduce observations, caution should be taken if using them to interpret the behaviour of real galaxy populations.

\subsubsection{AGN feedback}
\label{sec:AGN_model}
In \eagle \, simulations (except for the \NoAGN \, model), feedback from AGN quenches star formation in massive galaxies, shapes the gas profiles in the inner parts of their host haloes, and regulates the growth of BH. When the host halo mass of a given galaxy increases above $10^{10}~h^{-1}~\rm{M}_\odot$, a seed black hole of mass $10^5~h^{-1}~\rm{M}_\odot$ is placed inside it following \citet{Springel2005b}. The black hole then grows as a result of mergers and gas accretion. The model assumes a modified Bondi-Hoyle accretion rate (\citealt{RosasGuevara2015}; \citealt{Schaye2015}) that is regulated with a viscosity parameter $C_{\rm visc}$. AGN feedback is implemented thermally and stochastically, choosing random neighbouring particles and heating them with a temperature increment $\Delta T_{\rm AGN}$. A higher value of $\Delta T_{\rm AGN}$ implies more energetic and intermittent feedback events, leading to reduced radiative losses within the ISM. As shown by \citet{DeRossi2017}, a higher $\Delta T_{\rm AGN}$ drives a stronger AGN feedback impact on metallicity scaling relations of massive galaxies.

With the exception of the \NoAGN \, model, the \eagle \, simulations analysed here adopt a value of $C_{\rm visc}=2\pi$. In the reference model, $\Delta T_{\rm AGN}=10^{8.5}~\rm{K}$ is assumed, while, in the \AGNdT \, model (higher AGN feedback temperature increment), $\Delta T_{\rm AGN}=10^9~\rm{K}$. In the \NoAGN \, model, the BH growth and AGN feedback implementations are entirely turned off.

Finally, it is worth mentioning two caveats.  Throughout this study, we assess the characteristics of black holes associated with each galaxy by utilising the galaxy catalogues publicly available through the \eagle \, database. As mentioned in \cite{McAlpine2016}, the variable representing the black hole mass (denoted $M_{\rm BH}$, in this article) does not correspond to the mass of the central BH in a galaxy, but rather represents the cumulative value of all BH assigned to that subhalo.  Nonetheless, for  $M_{\rm BH} > 10^6~\rm{M}_\odot$, this closely approximates the mass of the most massive BH. Additionally, care should be taken when interpreting the variable used to quantify the accretion rate of BH ($\dot{M}_{\rm BH,acc}$, in this paper). The time sampling of the simulation outputs may not capture the high temporal variability of BH accretion rates accurately.

As previously indicated, Table~\ref{tab:simus} summarises the values of the main subgrid parameters implemented in the \eagle \, simulations used in this paper.

\subsection{Galaxy sample and definitions}
\label{sec:selection}
Following \citet{DeRossi2017}, we selected simulated galaxies with $M_\star \geqslant 10^9~\rm{M}_\odot$ to avoid resolution issues.  The number of systems resulting from this selection criteria in our different simulations is indicated in the last column of Table~\ref{tab:simus}.  Note that both, central and satellite galaxies are included in our galaxy samples.

As different feedback models predict galaxy populations with different galaxy size distributions (see, e.g. \citealt{Crain2015}), our main analysis is based on integrated galaxy properties calculated without imposing aperture limits (i.e. all particles identified as belonging to the galaxy by {\sc subfind} are considered). In this way, we can focus on the general effects of feedback on the global properties of galaxies and not on their spatial distribution. Thus, unless stated otherwise, we do not apply aperture corrections to galaxy properties. However, for the purpose of comparing our results with observations presented in \citetalias{Lara-Lopez2019} (Section~\ref{sec:comp_obs}), we recompute simulated quantities within similar apertures as those employed in their fiducial sample, to which we will refer to as `LL19 sample'. As shown in \citet{DeRossi2017} and \citetalias{Lara-Lopez2019}, aperture effects do not significantly affect the main features of $y_{\rm eff}$ and metallicity scaling relations, but can generate moderate changes in their slopes and absolute normalisations. 

In this work, the effective yield $y_{\rm eff}$ of a simulated galaxy is defined by equation~(\ref{eq:yeff}). Since gas-phase metal abundances are usually inferred from SF regions, unless otherwise specified and for the sake of consistency, we derive $y_{\rm eff}$, $\mu$ and $M_{\rm bar}$ considering only the SF gas component of our simulated galaxies. Besides, throughout this article, $M_{\rm gas}$ and $Z_{\rm gas}$ stand for the mass and metallicity of the SF gas component of a given galaxy, respectively. In addition, for the sake of comparison with previous works, we characterise the metallicity in terms of oxygen abundance (O/H), since oxygen is generally the most abundant heavy element in mass \citep{Maiolino2019}.  We estimate O/H as usual: ${\rm 12+log(O/H)=12+log(}N_{\rm O} / N_{\rm H})$, where $N_i$ represents the number density of the chemical element $i$ in the SF gas component of a galaxy. A comparison with similar quantities obtained from the total gas component is carried out in Section~\ref{sec:gas_heating}.

\section{Feedback effects on effective yields and associated 2D scaling relations}
\label{sec:2d_relations}
The effective yields of galaxies combine information regarding relevant properties of these systems, such as their gas and stellar content, and the corresponding metallicities. In this section, we study the effects of SN and AGN feedback on 2D fundamental scaling relations that involve such quantities. 
We try to determine how different feedback models affect $y_{\rm eff}$.

\begin{figure*}
    \includegraphics[width=2\columnwidth]{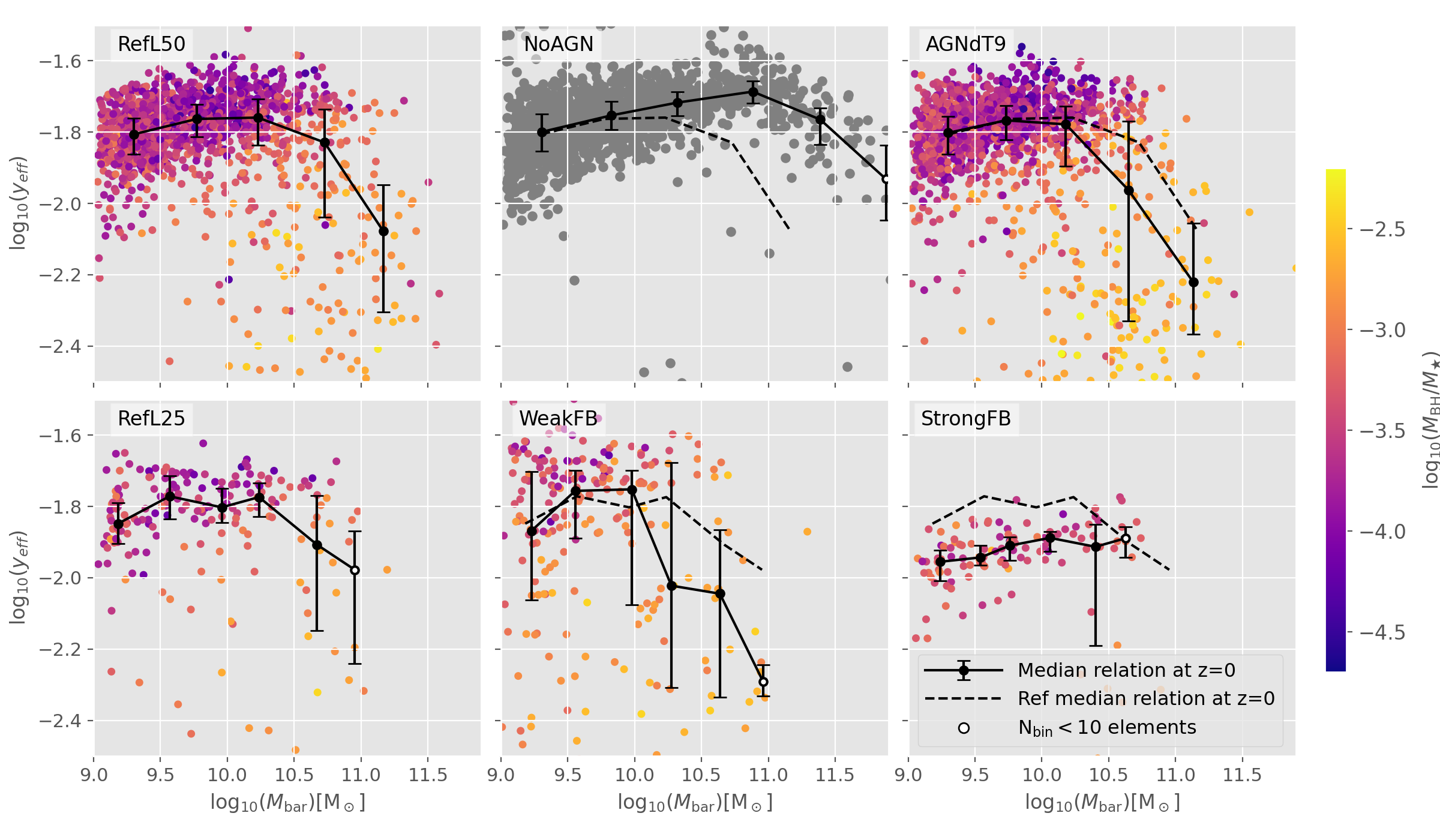}
    \caption{
    $M_{\rm bar} - y_{\rm eff}$ relation colour-coded according to the ratio between BH and stellar mass ($M_{\rm BH} / M_{\star}$) for $z=0$ \eagle \, galaxies. Top (bottom) panels show results for models with different AGN (SN) feedback parameters (see Table~\ref{tab:simus}, for details), as indicated in the legend. Solid lines represent the $z = 0$ median relation, dotted lines in middle and left panels depict the $z = 0$ median relation for the reference model. Error bars denote the 25th and 75th percentiles. Mass bins populated with less than 10 elements ($N_{\rm bin}<10$) are marked with white circles.
    }
    \label{fig:yeff_vs_Mbar}
\end{figure*}

\begin{figure*}
	\includegraphics[width=2\columnwidth]{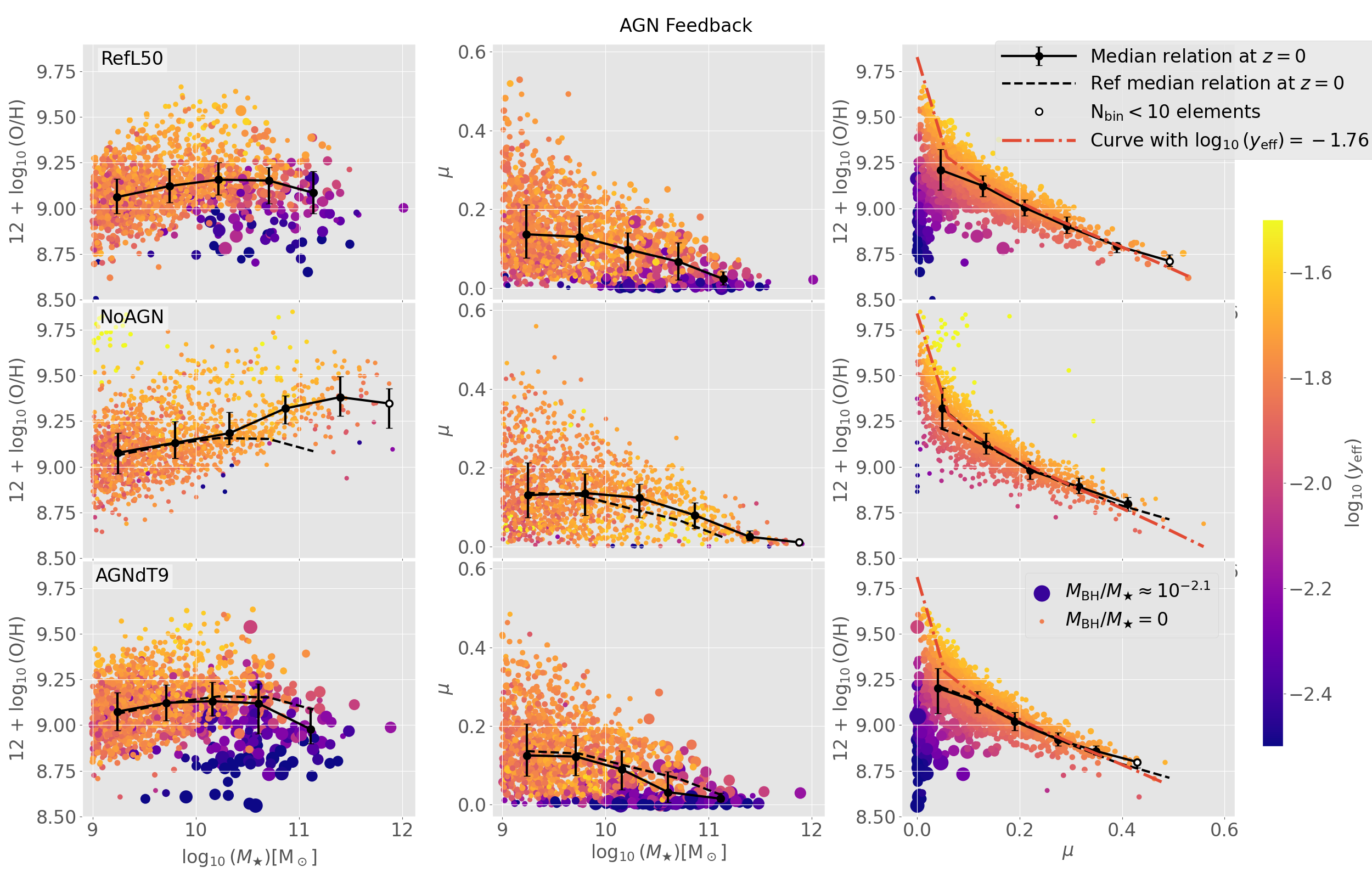}
    \caption{
    Simulations that evaluate the effects of varying the AGN feedback model (see Table~\ref{tab:simus}, for details). Left panels: $M_{\star} - {\rm O/H}$ relation. Middle panels: $M_{\star} - \mu$ relation. Right panels: $\mu - {\rm O/H}$ relation. The connection between each of these relations and the effective yields is highlighted by colour-coding symbols according to the values of $y_{\rm eff}$. Symbols are also scaled to the value of $M_{\rm BH}/M_\star$. Solid lines represent the $z = 0$ median relation, dotted lines in lower panels depict the $z = 0$ median relation for the reference model. Error bars denote the 25th and 75th percentiles. Mass bins populated with less than 10 elements ($N_{\rm bin} < 10$) are marked with white circles.
    }
    \label{fig:2D_AGN_models}
\end{figure*}

\subsection{Effective yield-baryonic mass relation}
\label{sec:yeff_Mbar_rel}
As mentioned in the Introduction, the well-known $y_{\rm eff} - M_{\rm bar}$ relation has been frequently used in the literature to evaluate the relative accumulated effects of feedback in observed galaxies of different masses.

In Fig.~\ref{fig:yeff_vs_Mbar}, we analyse the $y_{\rm eff} - M_{\rm bar}$ relation for our set of simulations. Symbols are colour-coded with $M_{\rm BH} / M_{\star}$, which quantifies the dominance of the BH in each galaxy. $M_{\rm BH} / M_{\star}$ can be regarded as a rough measure of the accumulated effects of AGN feedback during the evolution of a galaxy, in the sense that, at a fixed total stellar mass, galaxies with larger (smaller) $M_{\rm BH}$ are expected to have been more (less) significantly affected by AGN feedback along their formation histories.

According to the top panels in Fig.~\ref{fig:yeff_vs_Mbar} and consistently with results from \citetalias{Lara-Lopez2019}, variations of the AGN feedback model affect mostly $y_{\rm eff}$ of massive galaxies ($M_{\rm bar} \gtrsim 10^{10}~{\rm M}_{\odot}$). At low masses, where $M_{\rm BH} / M_{\star} \lesssim 10^{-3.5}$ for the bulk population, all simulations predict similar weak $y_{\rm eff} - M_{\rm bar}$ correlations with low scatter and high $y_{\rm eff}$. At higher $M_{\rm bar} \gtrsim 10^{10}~{\rm M}_{\odot}$, the \NoAGN \, model predicts the highest $y_{\rm eff}$ at a given mass.  On the other hand, the \RefAGN \, and \AGNdT \, models predict a strong median decrease of $y_{\rm eff}$ with $M_{\rm bar}$, caused by the presence of massive galaxies with dominant BH ($M_{\rm BH} / M_{\star} \gtrsim 10^{-3.5}$) and very  low $y_{\rm eff}$. The scatter at high masses also increases with $\Delta T_{\rm AGN}$. For the \RefAGN \, and \AGNdT \, simulations, the low $y_{\rm eff}$ of massive galaxies can be associated with the accumulated impact of AGN feedback driven by their dominant BH, which can heat SF gas, suppress star formation and chemical evolution, and drive outflows of metal-enriched material out of galaxies (see, \citealt{DeRossi2017} and \citetalias{Lara-Lopez2019}, for a discussion). Thus, our findings suggest a close connection between $y_{\rm eff}$ and $M_{\rm BH} / M_{\star}$.

The effects of varying the SN feedback efficiency are presented in the bottom panels of Fig.~\ref{fig:yeff_vs_Mbar}. In the case of the strong SN feedback model, all galaxies show low $M_{\rm BH} / M_{\star}$ and $y_{\rm eff}$, which are below those corresponding to systems of similar masses in the \RefSN \, model.  On the other hand, the \WeakFB \, model predicts a higher percentage of galaxies with high $M_{\rm BH} / M_{\star}$ at all masses.  At low masses, the \WeakFB \, and \RefSN \, models predict similar median  $y_{\rm eff} - M_{\rm bar}$ relations, whereas, at higher masses, the \WeakFB \, simulation shows a stronger $y_{\rm eff} - M_{\rm bar}$ anti-correlation, as a consequence of the higher number of galaxies with dominant BH. The \WeakFB \, model also predicts a larger scatter. Hence, it is important to highlight that the impact of SN feedback extends beyond the direct removal of metal-enriched material, as observed in the \StrongFB \, simulation. SN feedback also appears to affect $y_{\rm eff}$ through the regulation of the BH-to-stellar mass growth: interestingly, a weaker SN feedback tends to drive a more pronounced impact from AGN feedback, whereas a stronger SN feedback diminishes the relevance of AGN feedback. As already discussed by \citet{Henriques2019}, SN feedback has a critical role in regulating the efficiency of AGN feedback; when the stellar mass reached by the galaxy is large enough to avoid mass ejection by SN feedback, more cold gas becomes available for star formation and BH growth, thus giving place to AGN feedback.

According to the aforementioned findings, both SN and AGN feedback seem to affect the shape, normalisation and scatter of the $y_{\rm eff} - M_{\rm bar}$ relation in different ways, which suggests that its features can provide clues about the relative accumulated effects of feedback in real galaxies of different masses. Considering the predictions of our simulations: 1) at a given $M_{\rm bar }\ga 10^{10}~{\rm M}_{\sun}$, lower $y_{\rm eff}$ can be associated with a stronger impact of AGN feedback; 2) at a given $M_{\rm bar }\la 10^{10}~{\rm M}_{\sun}$, galaxies with higher than the median $y_{\rm eff}$ can be associated with a weaker SN feedback efficiency {\em and} a weaker AGN feedback impact; 3) at a given $M_{\rm bar }\la 10^{10}~{\rm M}_{\sun}$, galaxies with $y_{\rm eff}$ well below the median, correspond generally to systems with a strong AGN impact and a weak SN feedback efficiency.

Finally, we note that some previous works that analyse {\sc EAGLE} galaxies in the context of analytical models adopt a `true stellar yield' parameter $y_{Z} = 0.04$ (${\log}_{10}(y_{Z})\approx -1.4$; e.g. \citealt{Sharma2020}; \citealt{Zenocratti2022}), which seems to be suitable for describing the behaviour of most simulated systems.   Such value is consistent with the net yields expected for simple stellar populations (SSPs) with a Chabrier IMF \citep[e.g.][]{Madau2014}.  The {\em effective} yields shown in Fig.~\ref{fig:yeff_vs_Mbar} are below the aforementioned {\em true} yield parameter, a fact that is expected as simulated galaxies are not closed boxes.  
In addition to SN and AGN feedback, they are affected by the accumulated effects of past mergers  and gas flows.  Besides, simulated galaxies are composed by a mix of SSPs with different metallicities and, thus, different net yields.

\subsection{Metallicities, stellar masses and gas fractions}
\begin{figure*}
	\includegraphics[width=2\columnwidth]{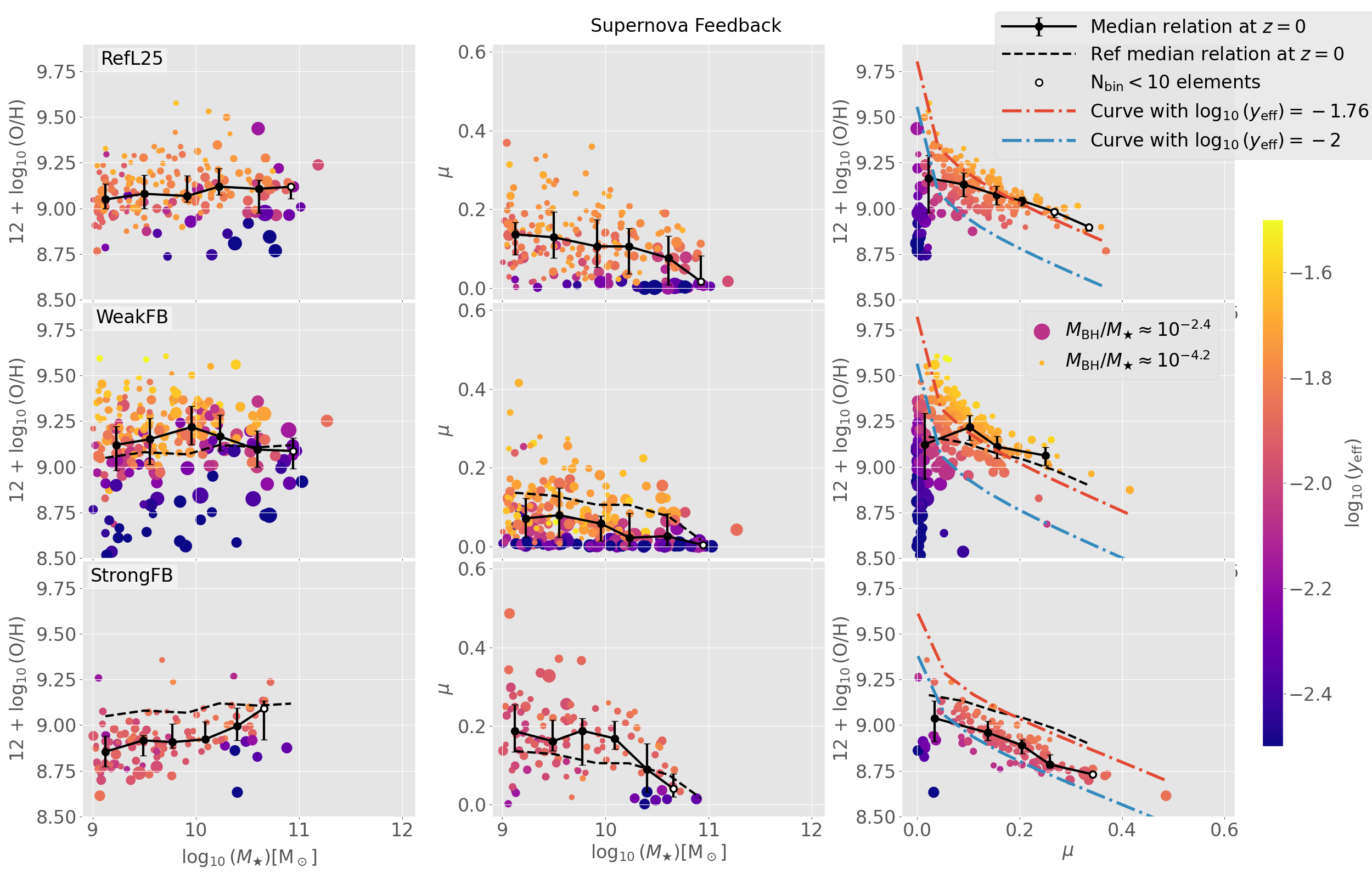}
    \caption{
    Simulations that evaluate the effects of varying supernova (SN) efficiency (see Table~\ref{tab:simus}, for details). Left panels: $M_{\star} - {\rm O/H}$ relation. Middle panels: $M_{\star} - \mu$ relation. Right panels: $\mu - {\rm O/H}$ relation. The connection between each of these relations and the effective yields is highlighted by colour-coding symbols according to the values of $y_{\rm eff}$. Symbols are also scaled to the value of $M_{\rm BH}/M_\star$. Solid lines represent the z = 0 median relation, dotted lines in lower panels depict the $z = 0$ median relation for the reference model. Error bars denote the 25th and 75th percentiles. Mass bins populated with less than 10 elements (N$_{\rm bin} < 10$) are marked with white circles.
    }
    \label{fig:2D_SN_models}
\end{figure*}

As effective yields are calculated from  metallicities, stellar masses and gas fractions (equation~\ref{eq:yeff}), the study of feedback effects on such quantities is relevant to explain their impact on $y_{\rm eff}$. In this section, we start exploring well-known 2D galaxy scaling relations involving ${\rm O/H}$, $M_\star$ and $\mu$. 
Consequences of varying AGN and SN feedback models on such relations are analysed in Fig.~\ref{fig:2D_AGN_models} and Fig.~\ref{fig:2D_SN_models}, respectively, where scatter plots are colour-coded by $y_{\rm eff}$, and symbols are scaled to the value of $M_{\mathrm{BH}}/M_{\star}$.

\subsubsection{AGN feedback effects}
Left panels of Fig.~\ref{fig:2D_AGN_models} show the M$_\star$Z$_{\rm g}$R for different AGN feedback models. At the low-mass end, different feedback prescriptions predict high $y_{\rm eff}$ and a slightly similar positive M$_\star$Z$_{\rm g}$R slope. We notice a break-point at $M_\star \sim 10^{10.3}{\rm M}_\odot$, where the \RefAGN \, and \AGNdT \, models predict a transition from a correlation to an anti-correlation for higher masses.  On the other hand, a steeper correlation is obtained for the \NoAGN \, model at such intermediate masses, with a gradual flattening towards $M_{\star} \gtrsim 10^{11.5}~{\rm M}_{\odot}$. The steeper negative slope of the high-mass end of the M$_\star$Z$_{\rm g}$R with increasing $\Delta T_{\rm AGN}$ is accompanied by an increasing number of metal-poor galaxies with low $y_{\rm eff}$. In fact, comparing the three simulations, galaxies with the lowest $y_{\rm eff}$ are obtained in the \RefAGN \, and \AGNdT \, implementations, and correspond to systems with high $M_\star$ and low ${\rm O/H}$. In addition, the scatter plots symbols in Fig.~\ref{fig:2D_AGN_models} are scaled with $M_{\rm BH} / M_\star$. It is clear that massive galaxies with lower metallicities have more dominant BH, indicating that they could have been more affected by AGN feedback. On the contrary, we note that the highest values of $y_{\rm eff}$ are obtained in the simulation \NoAGN \, and correspond to systems with low $M_\star$, high ${\rm O/H}$ and null $M_{\rm BH}/ M_\star$. Our findings are in agreement with \citet{DeRossi2017}, who claimed that the negative slope of the M$_\star$Z$_{\rm g}$R at high $M_\star$ can be driven by AGN feedback through two different main channels: i) the heating of SF gas and its subsequent change to a non-star-forming (NSF) phase, which drives the quenching of star formation and, consequently, the suppression of chemical evolution; ii) the ejection of metal-enriched gas out of galaxies. $y_{\rm eff}$ seem to be good tracers of this situation, reaching lower and lower values as the global metallicity decreases due to the increasing AGN feedback influence.

As seen in the middle panels of Fig.~\ref{fig:2D_AGN_models}, all AGN feedback models generate similar median $M_{\star} - \mu$ anti-correlations, with $\mu$ ranging between $0$ and $0.5$ for smaller galaxies, and showing almost negligible values, for massive ones. Although weaker, there is a trend for massive galaxies of decreasing their SF gas fraction with increasing $\Delta T_{\rm AGN}$. Such a behaviour is expected given that AGN feedback heats the gas, fostering its transition to a NSF phase. It is also worth noting that, as $\Delta T_{\rm AGN}$ increases, there is a higher number of galaxies with low $y_{\rm eff}$ and very low $\mu$, specially for high-mass galaxies. Therefore, considering equation~(\ref{eq:yeff}) and our previous analysis of the M$_\star$Z$_{\rm g}$R, the lower $y_{\rm eff}$ at the high-mass end are a consequence of both, the lower $\mu$ and lower metallicity of the corresponding galaxies.

Finally, right panels of Fig.~\ref{fig:2D_AGN_models} show the $\mu - {\rm O/H}$ relation. Consistently with the findings of \citet{DeRossi2017} for the \eagle \, `Recal' model, our simulations predict a tight decrease of metallicity with increasing SF gas fraction, showing a larger scatter towards lower values of $\mu$. Typical galaxies with $\mu \ga 0.1$ depict a median relation that follows roughly a curve corresponding to a constant $\log_{10} (y_{\rm eff}) \approx -1.76$. Interestingly, such tight relation represents well the behaviour of galaxies derived from our three different AGN feedback models, with the only exception of gas-poor systems. In particular, when AGN feedback is turned off, a higher number of gas-poor systems show higher metallicities and $y_{\rm eff}$.  On the other hand, the \RefAGN \, model predicts a significant population of gas-poor systems with lower than average metallicity and lower than average $y_{\rm eff}$.  And, the latter population is even larger for the \AGNdT \, model.  As shown before, very low $y_{\rm eff}$ are related to the presence of galaxies with dominant BH (i.e. high $M_{\rm BH}/M_{\star}$ values) and, hence, probably more affected by AGN feedback. In particular, three ranges of $y_{\rm eff}$ are clearly distinguished in the $\mu - {\rm O/H}$ plane: 1) galaxies with high $y_{\rm eff}$ follow the $\log_{10}(y_{\rm eff})\approx-1.76$-curve, showing slightly higher than average metallicity at a given $\mu$; 2) galaxies with intermediate $y_{\rm eff}$ also follows the same curve, showing slightly lower than average metallicity at a given $\mu$; 3) low $y_{\rm eff}$ correspond to galaxies with low metallicities and low gas-fractions, significantly departing from the $\log_{10}(y_{\rm eff}) \approx -1.76$-curve. The latter region is only significantly populated when AGN feedback is turned on and correspond to massive galaxies ($M_\star \ga 10^{10}~\rm{M}_\odot$) with the most dominant BH.

\subsubsection{SN feedback effects}
In this section, we explore the effects of varying SN feedback efficiency on ${\rm O/H}$, $M_\star$ and $\mu$.  Given the smaller box of the simulations analysed here (\RefSN, \WeakFB, \StrongFB; see Section~\ref{sec:simulations}), the number of galaxies in our selected samples is smaller than those studied in the previous section (\RefAGN, \NoAGN, \AGNdT ; see Table~\ref{tab:simus}).  We also note that, in the case of the \StrongFB \, simulation, the number of galaxies is a factor of $\approx 2$ lower than for the \WeakFB \, run (Table~\ref{tab:simus}). This is  related to the stronger efficiency of feedback in the former case, which leads to a decrease in the star formation of galaxies due to gas heating and SN-driven winds. These effects prevent that many galaxies surpass our imposed lower mass limit of $M_{\star} = 10^9~{\rm M}_{\odot}$.

In Fig.~\ref{fig:2D_SN_models}, left panels, the effects of different SN feedback prescriptions on the M$_\star$Z$_{\rm g}$R are compared. We clearly see distinct patterns for the different simulations.  As reported in previous works, the \RefSN \, model predicts a flat median relation with moderate scatter.  Regarding the \WeakFB \, model, it leads to a higher normalisation for the median M$_\star$Z$_{\rm g}$R, which shows slightly higher positive (negative) slope at low (high) $M_{\star}$.  The \WeakFB \, model also produces a larger scatter in metallicity at all masses.  Such scatter is mostly generated by the appearance of a population of galaxies with low metallicities, low $y_{\rm eff}$ and dominant BH (larger symbols, see also Fig.~\ref{fig:yeff_vs_Mbar}). But, in contrast with \RefAGN \,  and \AGNdT \, simulations (see Fig.~\ref{fig:2D_SN_models}),  a significant number of dominant BH are located in low-mass galaxies in the \WeakFB \, simulation.  This could be caused by the weaker SN efficiencies in the latter case, which allows more gas to cool down and reach the galaxy centre, enhancing the growth of the central BH (see, also, the discussion in Section~\ref{sec:yeff_Mbar_rel}). According to these results, a weak feedback efficiency can lead to significant AGN feedback impact in galaxies of all masses, contributing to lower the metallicity even for low-mass systems.  On the other hand, the \StrongFB \, model do not lead to the formation of dominant BH, driving a M$_\star$Z$_{\rm g}$R with a positive slope along the whole mass range.  These trends support again a scenario where dominant BH are required for reproducing the flattening of the M$_\star$Z$_{\rm g}$R at high masses (\citealt{DeRossi2017}). In addition, the normalisation of the M$_\star$Z$_{\rm g}$R is lower for the \StrongFB \, model, which is expected as a stronger SN feedback fosters the ejection of metal enriched material out of galaxies.  

In the middle panels of Fig.~\ref{fig:2D_SN_models}, we compare the $\mu - M_{\star}$ relation for our three SN feedback scenarios. In the case of the reference model, $\mu$ decreases with increasing $M_{\star}$, with a large scatter around the median relation. Low values of $y_{\rm eff}$ can be associated with gas-poor galaxies. A similar behaviour is obtained for the \WeakFB \, model but, with a lower normalisation for $\mu$, and a higher number of gas-poor galaxies with low $y_{\rm eff}$. This is consistent with a more efficient SF process and gas consumption in the \WeakFB \, simulation: given the less significant influence of SN feedback, galaxies tend to form their stellar component earlier. In addition, as previously discussed, low-$y_{\rm eff}$ galaxies in these simulations also have dominant BH, which prevent further gas cooling, quenching star formation and chemical evolution. It is interesting to see that only systems with dominant BH reach negligible $\mu$. In the case of the \StrongFB \, simulations and contrary to expectations, low-mass galaxies show higher median $\mu$ than in the other simulations. This can be due to a stronger SN feedback heating during the peak of the SF histories of these systems, which could caused the ejection of gas from the shallower potential well of these galaxies, with the ejected material remaining in the outer part of them. As the potential wells become deeper at later epochs, the ejected material could have been re-accreted, driving an increase in $\mu$ and a decrease in metallicity; this explains the presence of a more significant population of galaxies with low ${\rm O/H}$ and high $\mu$ in the \StrongFB \, simulation. 
 
With respect to the $\mu - {\rm O/H}$ relation (Fig.~\ref{fig:2D_SN_models}, right panels), the \RefSN \, model predicts similar trends to those discussed before for the \RefAGN \, simulation: ${\rm O/H}$ decreases with increasing $\mu$, a larger scatter is obtained at lower $\mu$ and, once more, the median $\mu - {\rm O/H}$ relation follows a curve of constant $\log_{10}(y_{\rm eff}) \approx -1.76$. But, in contrast to the results obtained from variations of $\Delta T_{\rm AGN}$, variations in SN feedback efficiencies generate more significant departures from the $\log_{10}(y_{\rm eff}) \approx -1.76$ curve. For the \WeakFB \, model, the $\mu - {\rm O/H}$ relation is flatter, shows larger scatter and exhibits a small offset towards higher metallicities (and, hence, higher $y_{\rm eff}$)  with respect to the \RefSN \, model. On the other hand, for the \StrongFB \, model, the $\mu - {\rm O/H}$ relation presents a similar slope to that associated to the \RefSN \, model, but departs around $0.2$ dex towards lower metallicities (following roughly the curve of constant $\log_{10}(y_{\rm eff}) \sim - 2)$.

\subsection{Effective yield-BH connection}

\begin{figure*}
	\includegraphics[width=2\columnwidth]{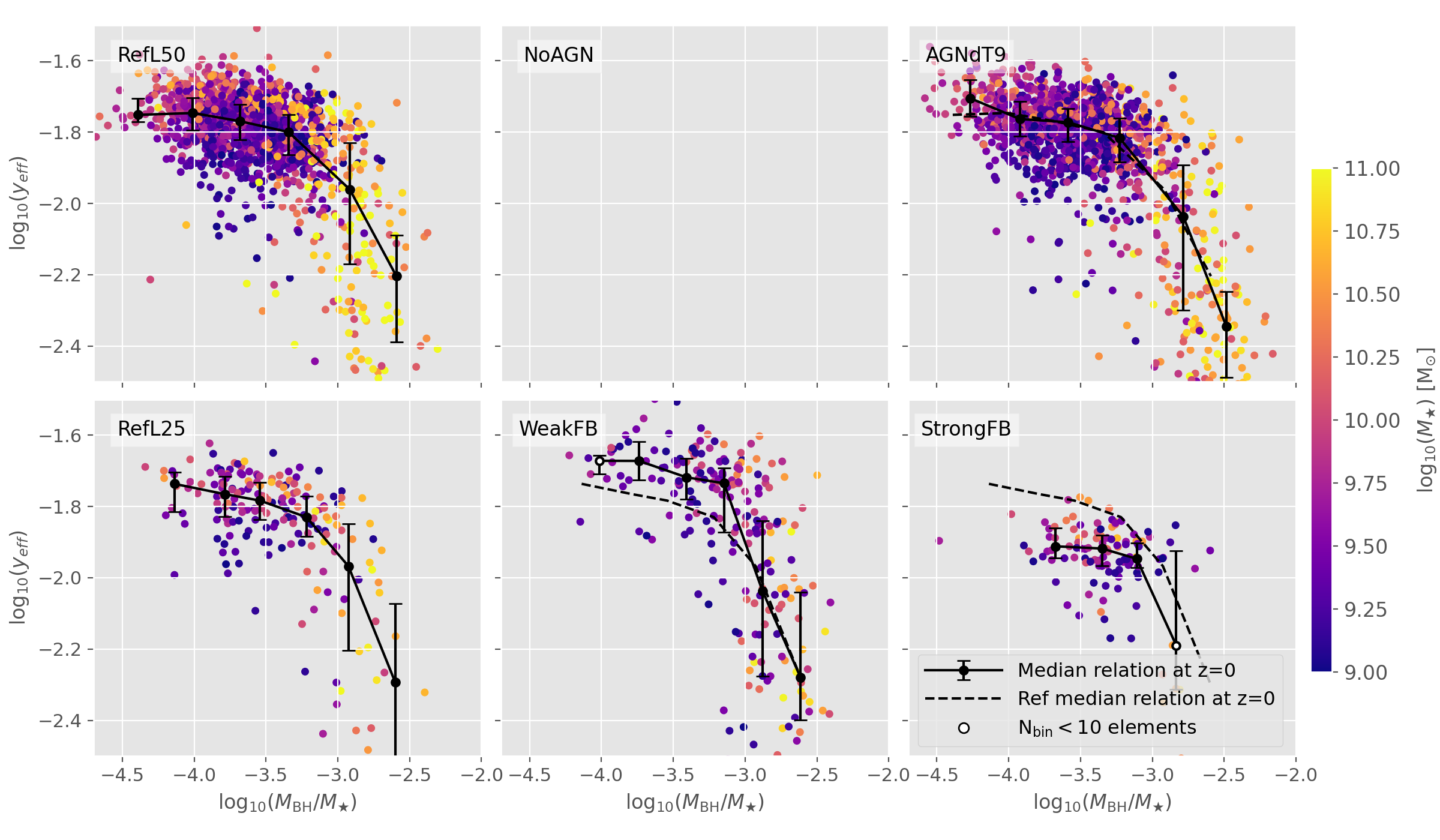}
    \caption{
    $y_{\rm eff}$ vs $M_{\rm BH}/M_{\star}$ for different AGN (top panels) and SN (bottom panels) feedback models (see Table~\ref{tab:simus}, for details). Symbols are colour-coded according to $M_{\star}$. Solid curves correspond to the median relations and error bars depict the 25th and 75th percentiles. For the sake of comparison, the median relation corresponding to the reference model for each simulation set is shown in each panel.
    }
    \label{fig:yeff_vs_MbhMstar}
\end{figure*}

The results obtained so far suggest the existence of an anti-correlation between $y_{\rm eff}$ and $M_{\rm BH}/M_{\star}$.  This is verified in Fig.~\ref{fig:yeff_vs_MbhMstar}, which presents a scatter plot of the dependence of $y_{\rm eff}$ on $M_{\rm BH}/M_{\star}$ for all models considered. Note that symbols are colour-coded according to $M_{\star}$. 

It is clear that, for galaxies with high $M_{\rm BH}/M_{\star} \gtrsim 10^{-3.5}$, all simulations predict a strong anti-correlation between  $y_{\rm eff}$ and $M_{\rm BH}/M_{\star}$. Although the median $y_{\rm eff} - M_{\rm BH}/M_{\star}$ relation seems not to depend on the feedback model, the number of galaxies with higher $M_{\rm BH}/M_{\star}$ increases for higher AGN heating temperatures. Besides, galaxies with $M_{\rm BH}/M_{\star} \gtrsim 10^{-3.5}$ tend to have $M_{\star} \gtrsim 10^{10}~{\rm M}_{\odot}$ in all simulations, with the only exception of the \WeakFB \, run. For the latter model, lower mass galaxies can have dominant BH, as we showed before. Our results indicate that AGN feedback seems to be the main responsible for the decrease of $y_{\rm eff}$ as $M_{\rm BH}/M_{\star}$ increases at $M_{\rm BH}/M_{\star} \ga 10^{-3.5}$. 

On the other hand, SN feedback has influence on the BH growth through the regulation of the gas reservoir. Clear signatures of SN feedback impact on $y_{\rm eff}$ are evident at low $M_{\rm BH}/M_{\star}$, where lower $M_{\star}$ and an almost constant $y_{\rm eff}$ value are obtained for all simulations. At $M_{\rm BH}/M_{\star} \lesssim 10^{-3.5}$, the \StrongFB \, and \WeakFB \, models predict a median $y_{\rm eff}$ of $\sim 10^{-1.9}$ and $\sim 10^{-1.7}$, respectively. On the other hand, the reference and \AGNdT \, models show, for similar $M_{\rm BH}/M_{\star}$, $y_{\rm eff} \sim 10^{-1.76}$, which corresponds to the median $y_{\rm eff}$ for the \NoAGN \, case. Additionally, it is interesting to note that the \StrongFB \, model predicts $y_{\rm eff}$ within a narrow range of intermediate  values ($\approx 10^{-2.2} - 10^{-1.8}$) for almost all galaxies, compared with the wider range of $y_{\rm eff}$ values covered by other feedback models ($\approx 10^{-2.4} - 10^{-1.6}$). Our trends could be explained considering that a strong SN feedback efficiency prevents from reaching very high $y_{\rm eff}$, whereas the lack of very massive BH (required for a strong AGN feedback impact), in the case of a strong SN feedback, prevents from reaching very low $y_{\rm eff}$. To sum up, Fig.~\ref{fig:yeff_vs_MbhMstar} suggests that, in the case of galaxies with no dominant BH, $y_{\rm eff}$ seem to be mostly determined by the efficiency of SN feedback, while, for galaxies with dominant BH, $y_{\rm eff}$ show a strong anti-correlation with $M_{\rm BH}/M_{\star}$ due, mainly, to AGN feedback.  

Finally, we note that the $M_{\rm BH}/M_\star$ range obtained from our analysis is consistent with recent observations of massive galaxies \citep[e.g.][]{Graham2023}.  This is not a surprise as the reference model in \eagle~simulations was calibrated using the relation between BH mass and stellar mass in a similar mass range (Section~\ref{sec:simulations}).

{\em Summarising the results from Section~\ref{sec:2d_relations}, AGN feedback tends to favour the simultaneous decrease of the SF gas fraction and metallicity of high-mass galaxies, leading to lower $y_{\rm eff}$. On the other hand, a stronger SN feedback efficiency leads to lower $y_{\rm eff}$ at all stellar masses as a consequence of the decrease of metallicity at a given SF gas fraction. And, a weak SN feedback efficiency drives a more complex behaviour, fostering the formation of dominant BH for a significant number of galaxies even at low $M_\star$: galaxies with dominant BH show lower ${\rm O/H}$, lower SF gas fractions and, hence, lower $y_{\rm eff}$, while galaxies with low $M_{\rm BH}/M_{\star}$ exhibit higher metallicities at a given SF gas fraction, and, hence, higher $y_{\rm eff}$.}

\section{The stellar mass-metallicity-gas mass parameter space}
\label{sec:3d_relations}

\begin{table*}
	\centering
	\caption{
    Output parameters ($C_1$, $C_2$ and $C_3$; fifth to seventh columns, respectively) from the fitting of equation~\eqref{eq:plane} to the \NoAGN \, simulated sample, as indicated in the {\em first column}. The 2D plane fit of the 3D data was performed using the R package {\sc hyper-fit} (\citealt{Hyperfit}). The {\em eighth column} indicates the intrinsic scatter orthogonal to the hyperplane ($\epsilon$). The {\em second column} indicates the stellar component used in the calculations: total stellar mass or stellar mass used in \citetalias{Lara-Lopez2019} (i.e. the component enclosed within a radius of 30 kpc, for simulations). The {\em third column} shows the considered gas component: total gas component, total SF gas phase or the gas component used in \citetalias{Lara-Lopez2019} (i.e. total hydrogen mass enclosed within a radius of 70 kpc, for simulations). The {\em fourth column} indicates the gas component used for estimates of ${\rm O/H}$: total gas component, total SF gas phase or that used in \citetalias{Lara-Lopez2019} (i.e. SF gas within a radius of 30 kpc, for simulations).
    }
	\label{table_fits}
	\begin{tabular}{rcccccccl} 
		\hline
		Data Sample & Stellar Mass & Gas Mass & Gas Metallicity & $C_1$ & $C_2$ & $C_3$ & $\epsilon$ & Plane identifier \\
		\hline
		NoAGN simulation & total & total SF & total SF & 0.844 ($\pm$ 0.013) & 2.49 ($\pm$ 0.05)&  -20.5 ($\pm$ 0.5) & 0.320 & \NoAGNfp\\
	    NoAGN simulation & total & total & total &  0.949 ($\pm$ 0.010) &  1.35 ($\pm$ 0.02) &  -11.0 ($\pm$ 0.2) & 0.248 & \NoAGNfp-tot-gas\\
	    NoAGN simulation & LL19 & LL19 & LL19 &  0.773 ($\pm$ 0.011) & 2.59 ($\pm$ 0.04) &  -21.1 ($\pm$ 0.4) & 0.299 & \NoAGNfp-LL19\\
		\hline
	\end{tabular}
\end{table*}

\begin{figure*}
	\includegraphics[width=1.6\columnwidth]{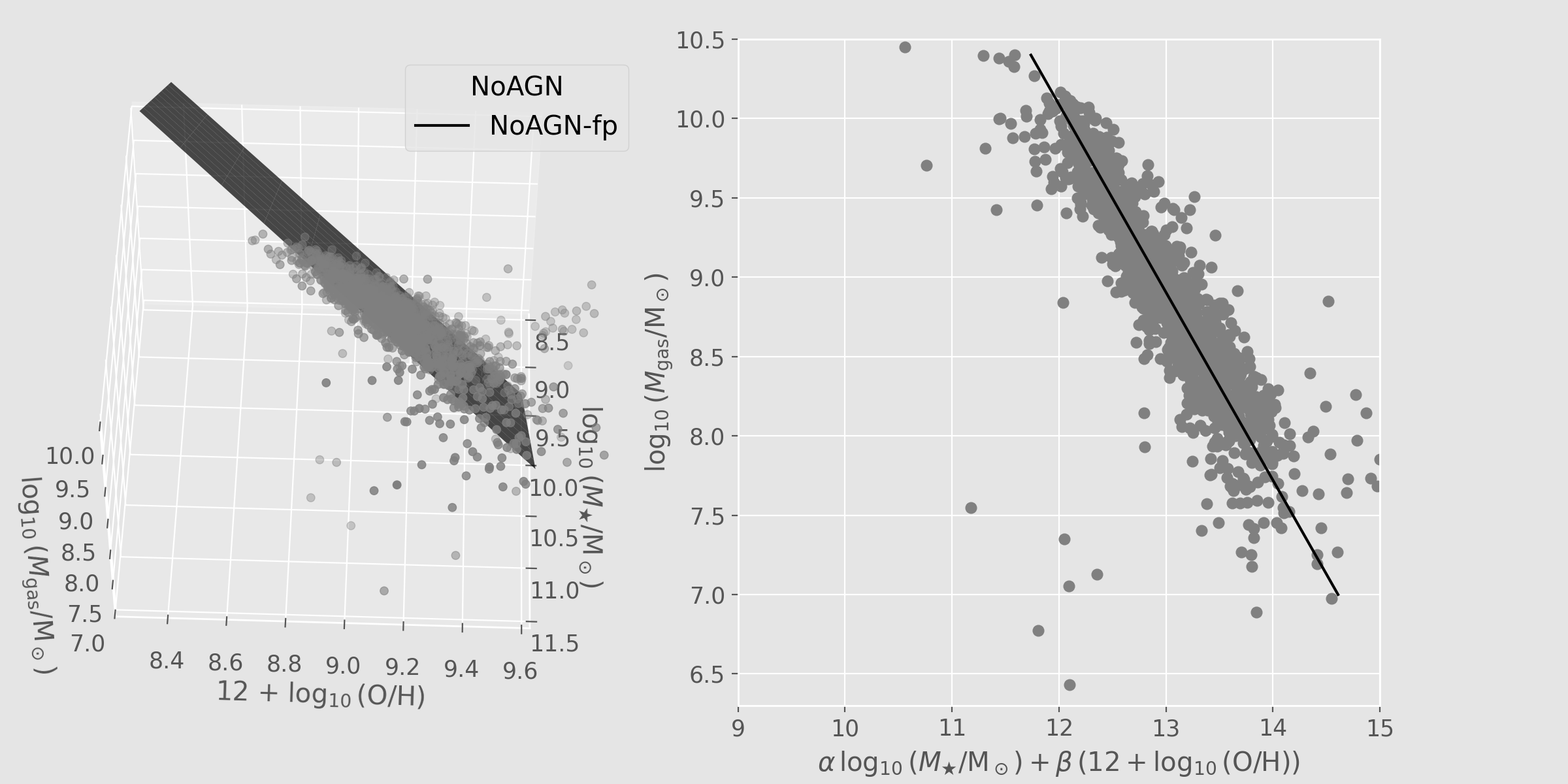}
	\includegraphics[width=1.6\columnwidth]{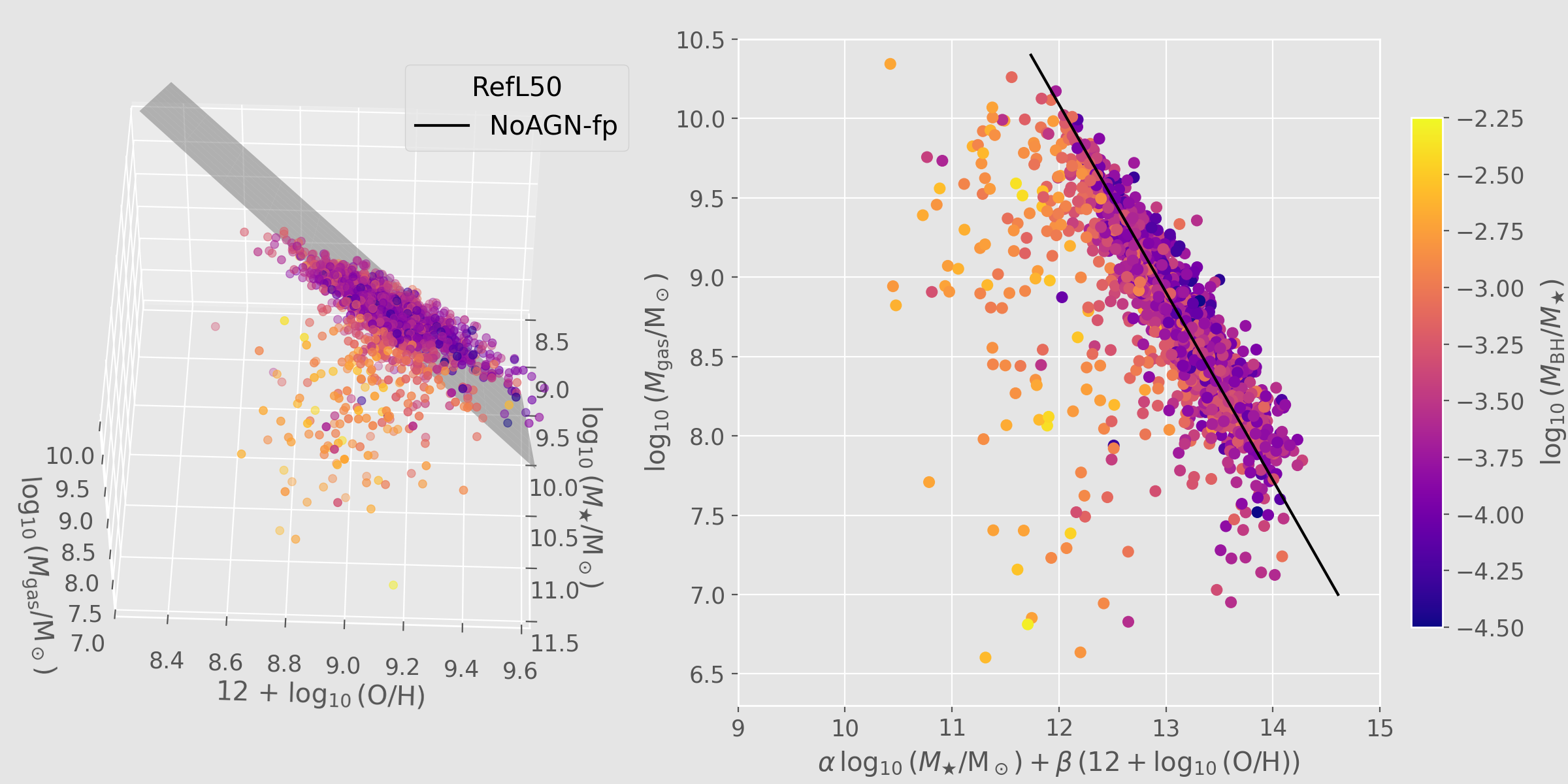}
	\includegraphics[width=1.6\columnwidth]{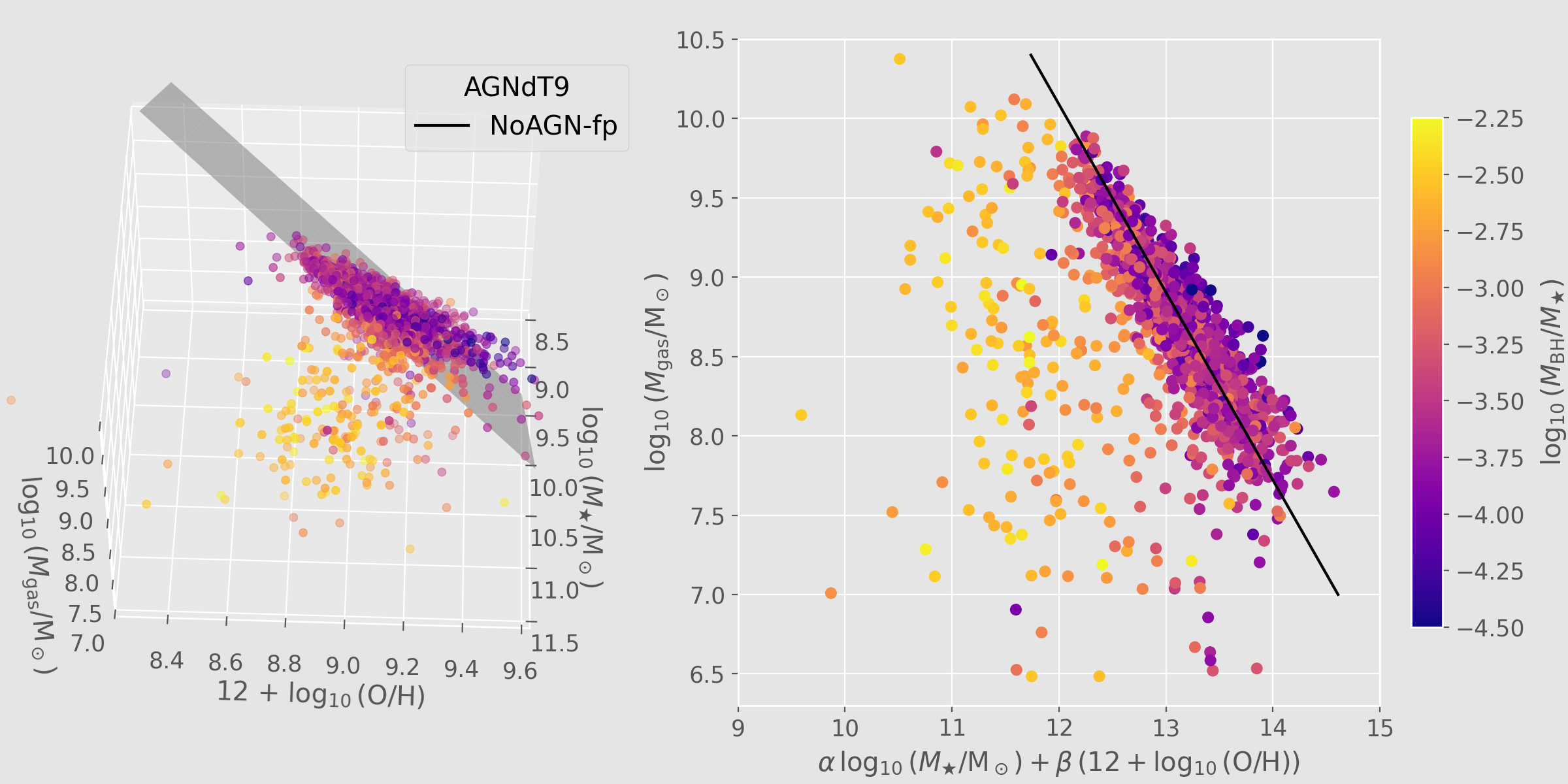}
    \caption{
    Scatter plot of galaxies in the $M_{\star} - {\rm O/H} - M_{\rm gas}$ space for simulations that evaluate the effects of varying AGN feedback (see Table~\ref{tab:simus}, for details). Left panel: 3D visualisation. Right panel: 2D projection with $\alpha = -1$ and $\beta = C_2 = 2.49$. Symbols are colour-coded with $M_{\rm BH} / M_{\star}$ as an indicator of the dominance of the BH mass in each galaxy.  Note that, for the \NoAGN \, simulation, $M_{\rm BH}$ is  null for all galaxies, so they are shown with an uniform grey colour. For more viewing angles, see the Supplementary Material.
    }
    \label{fig:3D_plot_AGN}
\end{figure*}

\begin{figure*}
        \includegraphics[width=1.6\columnwidth]{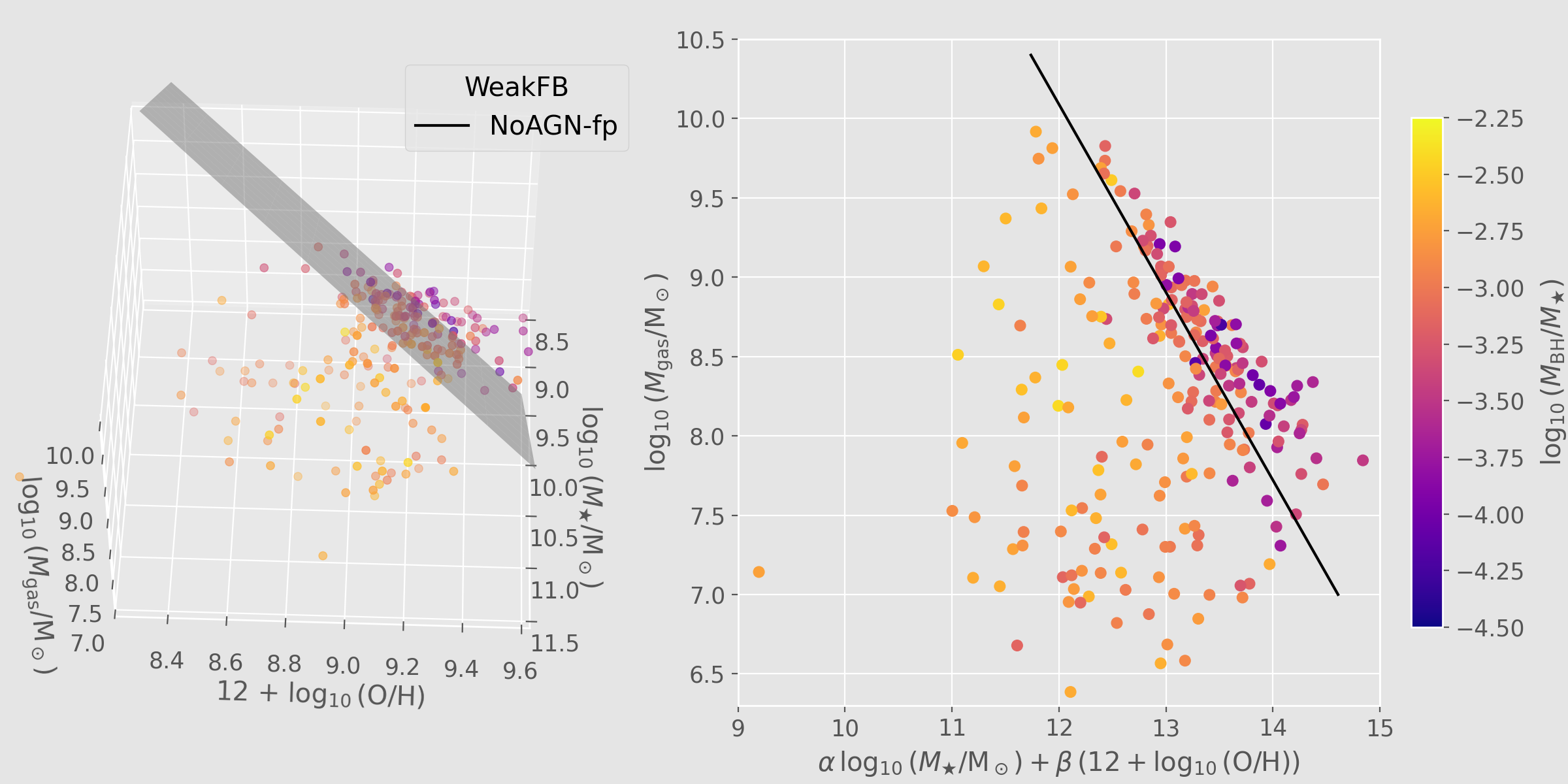}
	\includegraphics[width=1.6\columnwidth]{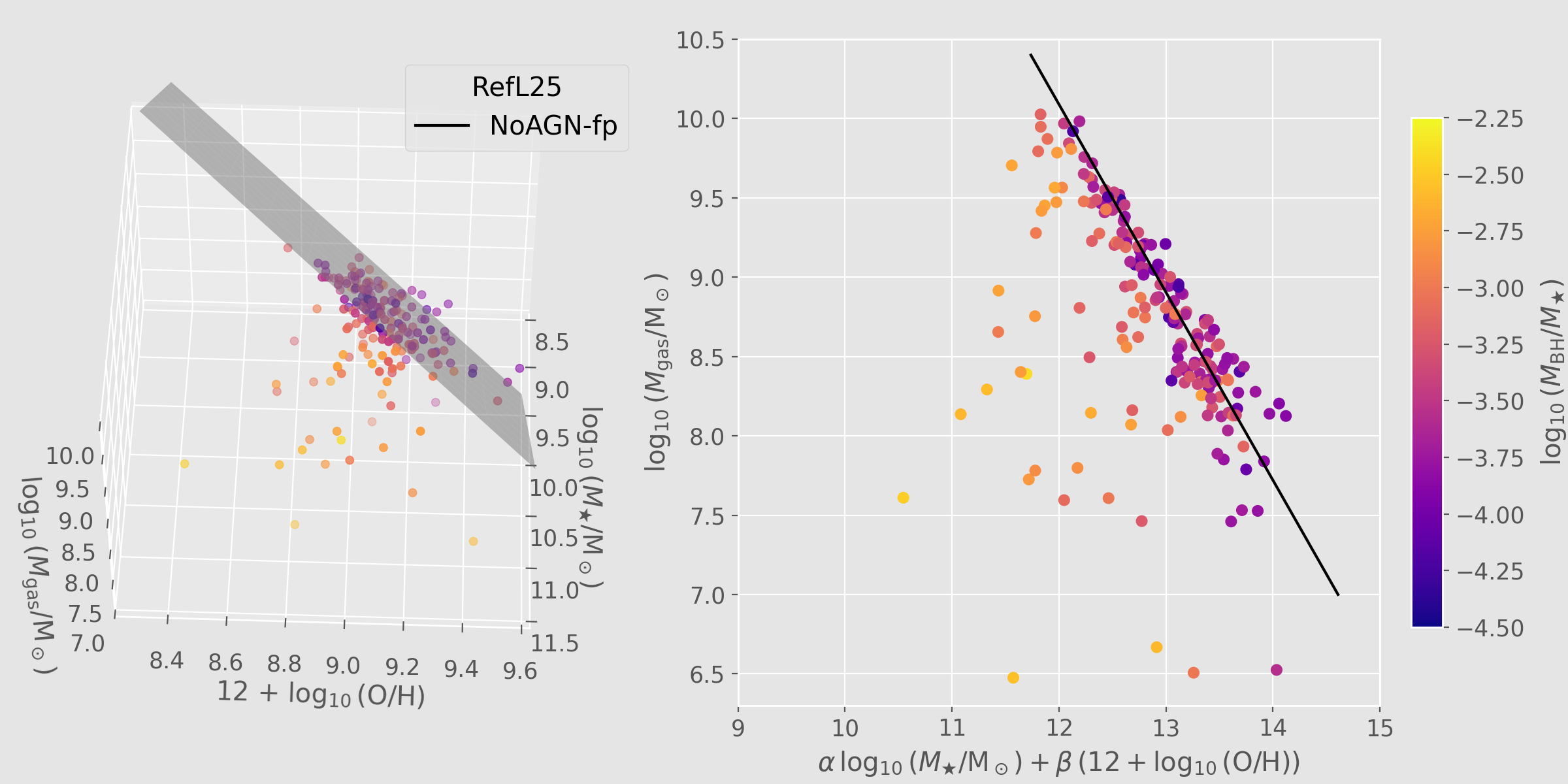}
	\includegraphics[width=1.6\columnwidth]{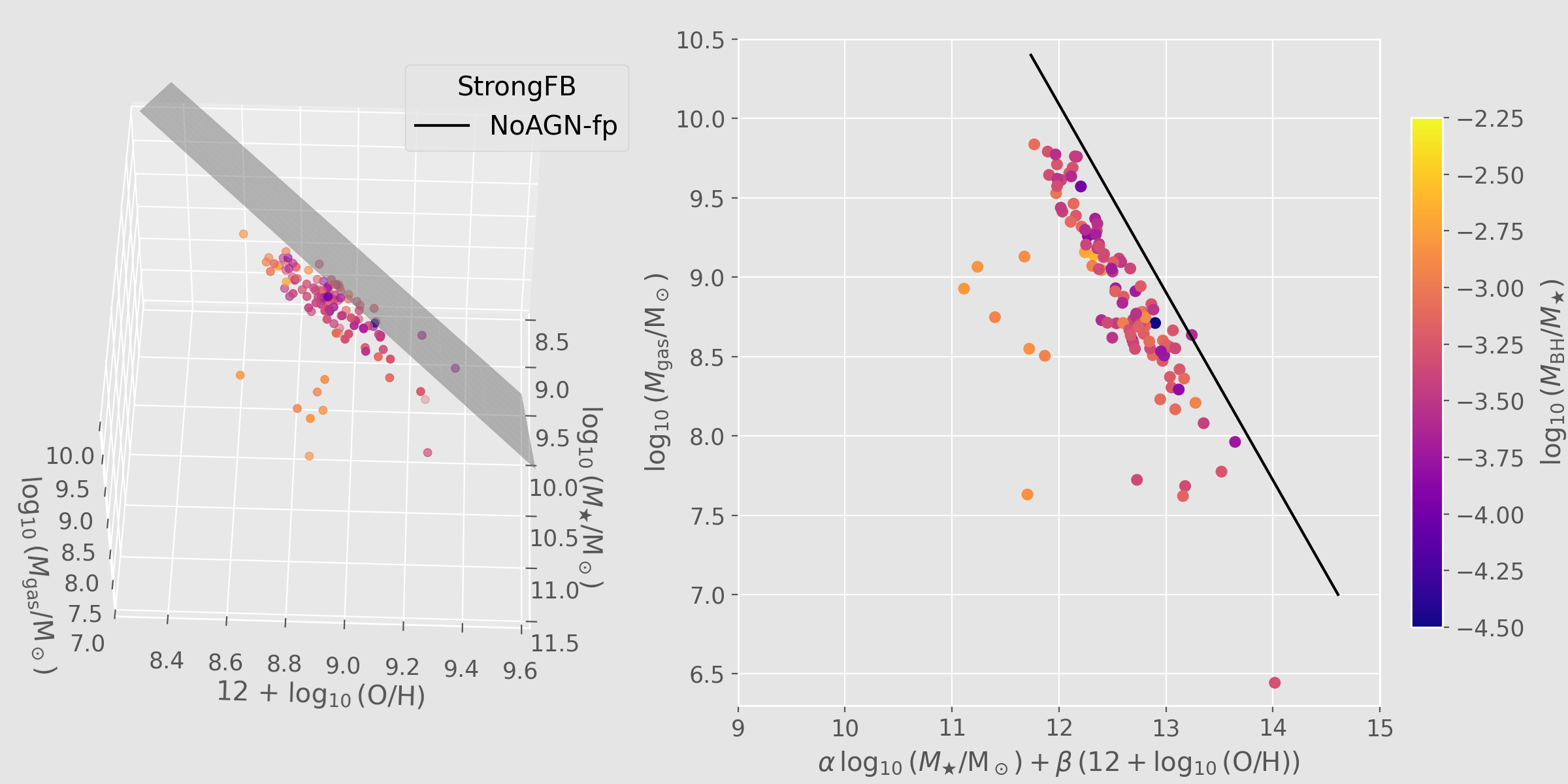}
 
    \caption{
    Similar to Fig.~\ref{fig:3D_plot_AGN}, but for the simulations that evaluate the effects of varying SN efficiency (see Table~\ref{tab:simus}, for details). For more viewing angles, see the Supplementary Material.
    }
    \label{fig:3D_plot_SN}
 \end{figure*}

\cite{Lara-Lopez2010} found a fundamental metallicity plane for SF galaxies, relating SFR, $M_\star$, and ${\rm O/H}$ (see also \citealt{Ellison2008,Mannucci2010}). Considering the results reported by \cite{Lara-Lopez2013}, such a plane could be a consequence of an underlying relation between $M_\star$, ${\rm O/H}$ and $\mu$ (see also \citealt{Bothwell2013} for another observational work and, e.g., \citealt{Lagos2016} and \citealt{DeRossi2017} for related analysis using \eagle \, simulations). As $y_{\rm eff}$ are defined from these key galaxy properties and taking into account the results discussed in the previous section, it is reasonable to expect that $y_{\rm eff}$ could trace the impact of feedback processes on the features of the aforementioned 3D galaxy metallicity scaling relations. In this section, we examine the relation between $M_\star$, ${\rm 12+\log_{10}({\rm O/H})}$ and $M_{\rm gas}$ (hereafter, M$_{\rm \star,g}$Z$_{\rm g}$R), for our complete set of \eagle \, simulations. We aim at evaluating how feedback processes affect the relations between these three fundamental properties and try to assess the role of $y_{\rm eff}$ as a feedback indicator.

Fig.~\ref{fig:3D_plot_AGN} and Fig.~\ref{fig:3D_plot_SN} show the M$_{\rm \star,g}$Z$_{\rm g}$R for simulations with different AGN and SN feedback prescriptions, respectively.  Different symbols are colour-coded according to $M_{\rm BH}/M_{\star}$. For the \NoAGN \, simulation, $M_{\rm BH}$ is always zero so galaxies are shown with an uniform grey colour. Comparing results from different panels, we see that \AGNdT \, and \WeakFB \, simulations present the largest dispersion. On the other hand, the smallest dispersion is obtained for the \NoAGN \, simulation, which, indeed, predicts an M$_{\rm \star,g}$Z$_{\rm g}$R that can be well represented by a plane. Such \NoAGN \, plane also seems to be a good representation of the bulk of galaxies with no dominant BH (i.e. low $M_{\rm BH}/M_{\star}$) in other simulations.

We determined the characteristic plane associated with the \NoAGN \, model by performing a least-square fit, of a two order polynomial, with the R hyperplane fitting package {\sc hyper-fit} (\citealt{Hyperfit}). The plane is well described by the following expression:
\begin{equation}
        \log_{10}{\left( \frac{M_\star} {{\rm M}_\odot} \right)} =  C_1 \log_{10}\left( \frac{M_{\rm gas}}{{\rm M}_\odot} \right) + C_2 \left[12 + \log_{10} ({\rm O/H})\right]  + C_3,
\label{eq:plane}        
\end{equation}
where the parameters $C_1$, $C_2$ and $C_3$ are shown in Table~\ref{table_fits} for different data samples. The first row in the table corresponds to the plane plotted in Fig.~\ref{fig:3D_plot_AGN} and Fig.~\ref{fig:3D_plot_SN}, and discussed here. We will refer to this plane as the {\em NoAGN fitting plane} (hereafter, `\NoAGNfp').

The departures (residuals) from the \NoAGNfp \, can be quantified by calculating the orthogonal deviation of each galaxy $i$ from such a plane as:
\begin{equation}
\begin{aligned}
        {\delta}_i = &  \frac{-\left[\log_{10}(M_\star)\right]_i + C_2  \left[12+\log_{10}({\rm O/H})\right]_i}{\sqrt{1+C_2^2 + C_3^2}} + \\
                 & + \frac{C_1\left[\log_{10}( M_{\rm gas})\right]_i + C_3 }{\sqrt{1+C_2^2 + C_3^2}},
\end{aligned}
    \label{eq:residuals}
    \end{equation}
where the subscript $i$ indicates quantities corresponding to the given galaxy $i$.

For the sake of clarity, the following convention will be adopted in this work: at a fixed $M_{\star}$ and ${\rm O/H}$, galaxies with higher (lower) $M_{\rm gas}$ than the value corresponding to the \NoAGNfp \, are assumed to be located `over' (`under') the plane, presenting positive (negative) deviations with respect to it.

A comparison between the \NoAGNfp \, and the features of the M$_{\rm \star,g}$Z$_{\rm g}$R for different feedback models is carried out in Section~\ref{sec:4.1}. In Section~\ref{sec:4.2}, we analyse the connection between the latter findings and the distribution of $y_{\rm eff}$ values for galaxy populations in different simulations.  In Section~\ref{sec:4.3}, we evaluate how different feedback scenarios can affect the deviations of the M$_{\rm \star,g}$Z$_{\rm g}$R from the \NoAGNfp.

\subsection{Feedback impact on the M$_{\rm \star,g}$Z$_{\rm g}$R}
\label{sec:4.1}

In Fig.~\ref{fig:3D_plot_AGN} and Fig.~\ref{fig:3D_plot_SN}, we compare the best fitting plane obtained for the \NoAGN \, simulation (grey surface) with the M$_{\rm \star,g}$Z$_{\rm g}$R associated to different feedback models. We see that the  \NoAGNfp \, can roughly describe the behaviour of the vast majority of galaxies with no dominant BH. In addition, when varying SN or AGN feedback prescriptions, clear distinct trends are obtained for the location of galaxies with respect to the \NoAGNfp. 

Fig.~\ref{fig:3D_plot_AGN} evaluates the impact of AGN feedback on the M$_{\rm \star,g}$Z$_{\rm g}$R. As previously discussed, the increase of $\Delta T_{\rm AGN}$ leads to a more significant number of galaxies with higher $M_{\rm BH}/M_{\star}$. At the same time, the \AGNdT \, model predicts the highest number of galaxies spread below the \NoAGNfp, displaying also the most substantial negative deviations. Another key feature is the dependence of the distance to the plane on the dominance of the BH, which is quantified by the parameter $M_{\rm BH}/M_{\star}$. Galaxies with lower $M_{\rm BH}/M_{\star}$ tend to be located closer to the plane, with almost null deviations with respect to it or even slightly positive ones. Conversely, systems with higher $M_{\rm BH}/M_{\star}$ show larger negative deviations. Therefore, our findings suggest that galaxies tend to deviate down from the \NoAGNfp \, as they are more affected by the heating of gas by AGN feedback.

The consequences of varying SN feedback are analysed in Fig.~\ref{fig:3D_plot_SN}. In principle, we can clearly see that a change in the SN feedback efficiency affects not only the scatter of the M$_{\rm \star,g}$Z$_{\rm g}$R but also the displacement of the bulk of the galaxy population with respect to the \NoAGNfp. For the \StrongFB \, model, all galaxies tend to be located below the \NoAGNfp. As discussed before, a stronger SN feedback leads to a moderate average decrease in O/H, at a given $\mu$, for most of the galaxies (Section~\ref{sec:2d_relations}), which explains their moderate displacement down the \NoAGNfp. For such galaxies, $M_{\rm BH}/M_{\star} \sim 10^{-4} - 10^{-3}$, which are values well below the maximum ones reached by galaxies in other simulations ($M_{\rm BH}/M_{\star} \sim 10^{-2.5} - 10^{-2}$). Thus, in the case of the \StrongFB \, model, larger deviations from the \NoAGNfp \, are not possible given that AGN feedback effects seem to be limited.  As SN feedback efficiency decreases  (see plots corresponding to the simulation \WeakFB), most of the galaxies move, on average, closer to the \NoAGNfp, some of them reaching positive deviations with respect to it. It is interesting to highlight that, in the \WeakFB \, simulation, galaxies above the \NoAGNfp \, (purple galaxies) are those characterised by low values of $M_{\rm BH}/M_{\star}$ ($\la 10^{-3.5}$). Thus, these galaxies more closely resemble a closed box model for two main reasons. Firstly, the low SN feedback efficiency implemented in this model ensures a weak effect of stellar winds and energy injection due to SN feedback. Secondly, the low values of $M_{\rm BH}/M_{\star}$ indicate that they should not have been strongly affected by gas outflows caused by AGN feedback. On the other hand, the \WeakFB \, model also predicts a significant number of galaxies with high $M_{\rm BH}/M_{\star}$, which tend to be located at larger distances below the \NoAGNfp . As mentioned before, SN and AGN feedback processes seem to be indirectly intertwined for such galaxies: due to their lower SN feedback efficiency, they could have evolved faster due to an early 
efficient consumption of their cold gas reservoir, reaching $z=0$ with higher $M_{\rm BH}/M_{\star}$ and, thus, having been more affected by AGN feedback. 

A comprehensive analysis of the differences between galaxies in proximity to the \NoAGNfp \, and those situated at greater distances is undertaken in greater depth within Section~\ref{sec:nature_deviations}.

\subsection{Effective yields as feedback tracers}
\label{sec:4.2}
To evaluate the connection between $y_{\rm eff}$ and the feedback-driven features of the M$_{\rm \star,g}$Z$_{\rm g}$R in Fig.~\ref{fig:3D_plot_AGN} and Fig.~\ref{fig:3D_plot_SN}, we can compare the M$_{\rm \star,g}$Z$_{\rm g}$R predicted by different feedback models with the $y_{\rm eff}$ distribution of simulated galaxies. In this sense, we note that, since $y_{\rm eff}$ is given by gas metallicity and gas mass fraction (equation~\ref{eq:yeff}), a unique $y_{\rm eff}$ value is associated to each point within the 3D parameter space defined by $M_{\star}$, ${\rm O/H}$ and $M_{\rm gas}$. Interestingly, simulated galaxies located around the \NoAGNfp \, show a roughly constant $y_{\rm eff} \approx 10^{-1.76}$, being these trends consistent with the flatter $M_{\rm bar} - y_{\rm eff}$ relation described by galaxies in the \NoAGN \, simulation (see Fig.~\ref{fig:yeff_vs_Mbar}). Such characteristic value $y_{\rm eff} \approx 10^{-1.76}$ approximates the average $y_{\rm eff}$ of galaxies in the \NoAGN \, model. In addition, the bulk of the galaxy population in all simulations is located within a region where the \NoAGNfp \, seems to be a good local representation of a surface defined by $y_{\rm eff} \approx 10^{-1.76}$. In this context, orthogonal deviations from the \NoAGNfp \, in our 3D parameter space can be associated, locally, with variations in the values of $y_{\rm eff}$.  
As we will show in next section, the highest $y_{\rm eff}$ are reached by those galaxies with the largest positive deviations with respect to the plane, which correspond to systems with lower $M_{\rm BH}/M_{\star}$. On the other hand, galaxies that are located under the \NoAGNfp \, plane with larger deviations from it show higher $M_{\rm BH}/M_{\star}$ and lower $y_{\rm eff}$ (see, also, Fig.~\ref{fig:yeff_vs_MbhMstar}).

It is important to remember that our galaxy samples include both central and satellite galaxies of haloes (Section~\ref{sec:selection}). We verify that similar general trends would be obtained if only central galaxies were considered for our analysis; specially, a similar \NoAGNfp \, would emerge since few satellites lie within the considered mass range in the \NoAGN \, simulation. Satellite galaxies only tend to increase the dispersion towards lower masses as a consequence of the presence of systems with high $y_{\rm eff}$. This is consistent with the enhanced metallicity of \eagle \, satellite galaxies reported by \citet{Bahe2017}. These authors studied the \eagle \, reference model and concluded that satellite galaxies tend to be more metal enriched than equally massive centrals. They demonstrated that this phenomenon is primarily driven by the removal of metal-poor SF gas from the outer regions of galaxies, accompanied by the suppression of metal-poor inflows resulting from the removal of gas from the galaxy halo. Consequently, satellite galaxies in \eagle \, simulations present higher metallicities than central galaxies of the same mass, showing positive deviations from the \NoAGNfp.

\subsection{Residual Analysis}
\label{sec:4.3}

\begin{figure*}
    \includegraphics[width=2\columnwidth]{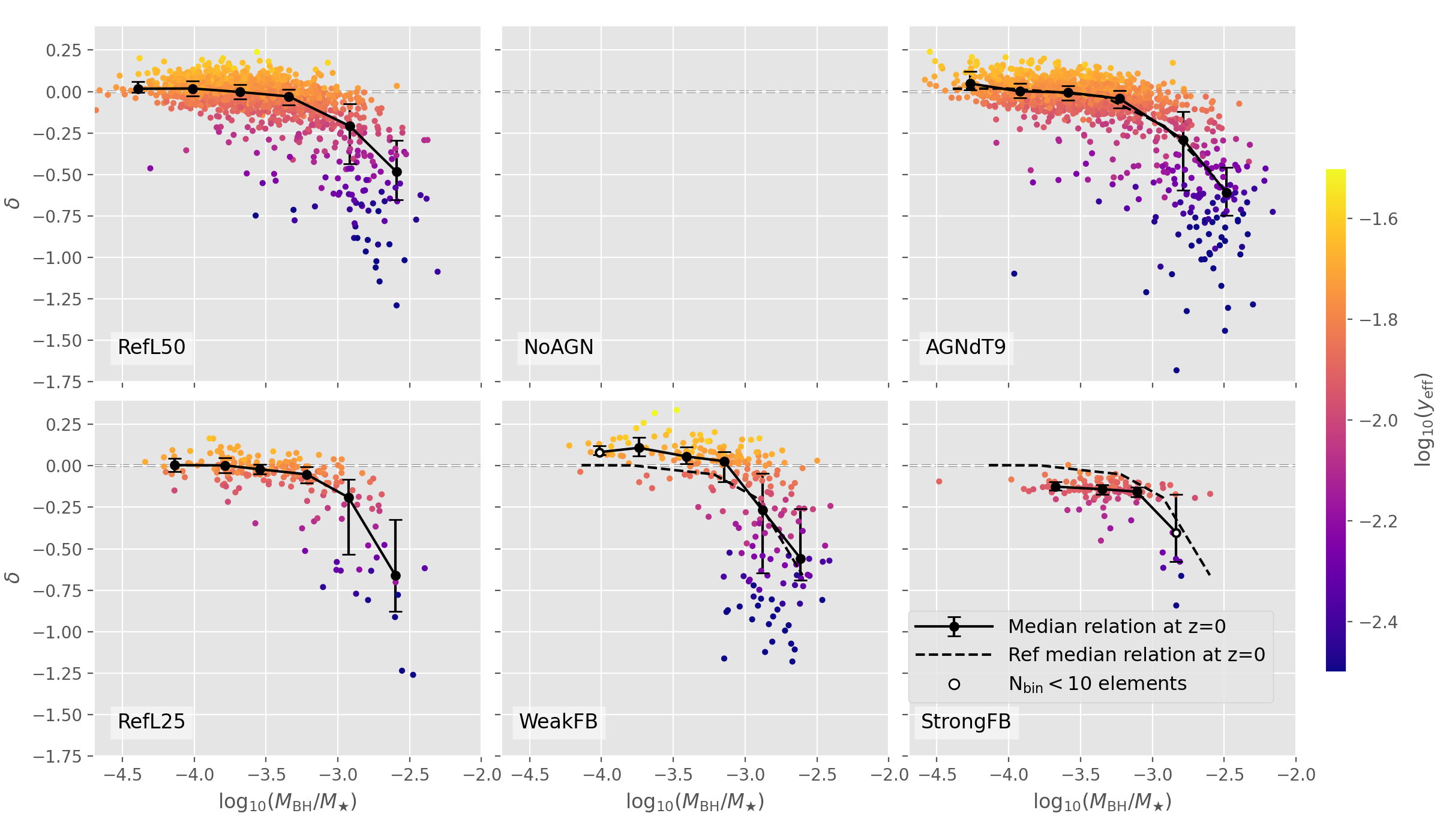}
    \caption{
    Residuals from the \NoAGNfp , $\delta$, vs $\log_{10} (M_{\rm BH} / M_{\star})$, colour-coded with $y_{\rm eff}$. Top (bottom) panels compare different AGN (SN) feedback models (see Table~\ref{tab:simus}, for details). Horizontal dotted lines highlight the null distance to the \NoAGNfp. Note that for the \NoAGN \, model, $M_{\rm BH}$ is null, so no galaxies are plotted in this case.
    }
    \label{fig:residue_MbhMstar}
\end{figure*}

In this section, we quantify the deviations from the \NoAGNfp\, by using the residuals $\delta$ defined in equation~\eqref{eq:residuals}. We remind that, at a given $M_\star$ and metallicity, $\delta$ is taken to be positive (negative) for higher (lower) values of $M_{\rm gas}$ with respect to the \NoAGNfp. 

Fig.~\ref{fig:residue_MbhMstar} shows $\delta$ as a function of $M_{\rm BH}/M_{\star}$ for different feedback models. Symbols are colour-coded according to $y_{\rm eff}$. For models including the reference SN feedback efficiency (\RefSN, \RefAGN, \AGNdT),  galaxies with less dominant BH ($M_{\rm BH}/M_{\star} \la 10^{-3.5}$) show negligible median residuals.  This is expected since the \NoAGN \, feedback model, from which the \NoAGNfp\ is derived, adopts a reference SN feedback efficiency and considers no AGN feedback effects; hence, such a behaviour should be recovered for galaxies with low $M_{\rm BH}/M_{\star}$ in simulations with the default SN feedback model. On the other hand, for galaxies with no dominant BH, the residuals become negative (positive) when implementing a stronger (weaker) SN feedback efficiency relative to the reference model. In the case of galaxies with more dominant BH ($M_{\rm BH}/M_{\star} \gtrsim 10^{-3.5}$), $\delta$ decreases with $M_{\rm BH}/M_{\star}$ in all simulations, as it is also evident from Fig.~\ref{fig:3D_plot_AGN}-\ref{fig:3D_plot_SN}; this is a consequence of the increasing relevance of AGN feedback. 

The tight relation between $y_{\rm eff}$ and $\delta$ is clear in Fig.~\ref{fig:residue_MbhMstar} (compare it, also, with Fig.~\ref{fig:yeff_vs_MbhMstar}), which can be well described by a linear function, as we checked. By performing a linear least squares fit to the data obtained in the \NoAGN\,model, this relation takes the form:
\begin{equation}
\label{eq:yeff_delta_relation}
   \log_{10} (y_{\rm eff}) \approx (0.905\pm0.005)\, \delta - (1.772\pm0.001),
\end{equation}
being valid for galaxies in all simulations. As discussed before, the \NoAGNfp \, {\em locally} represents a surface of constant $y_{\rm eff}$ in our 3D parameter space. $\delta$ measures the length of an orthogonal vector from the plane to a given galaxy, so that this vector should be parallel to the gradient of the scalar field defined by $y_{\rm eff}$. Hence, increasing $\delta$, implies continuously crossing isosurfaces associated with different $y_{\rm eff}$ values.

\section{Discussion}
\label{sec:discussion}
Our results suggest a close connection between the accumulated effects of AGN and SN feedback, and the features of scaling relations involving stellar mass, metallicity and gas mass. In addition, given that $y_{\rm eff}$ is defined from those quantities, it seems to be a good tracer to constrain different feedback scenarios. In particular, we detect a characteristic plane in the 3D parameter space defined by the aforementioned properties, which is surrounded by galaxies with a reference SN feedback and a negligible impact of AGN feedback. Interestingly, the plane locally describes a surface of constant $y_{\rm eff}$, so that orthogonal departures from the plane can be directly associated to variations in $y_{\rm eff}$ (i.e. different $y_{\rm eff}$ isosurfaces are crossed when moving away from the plane). An increasing influence of SN or AGN feedback, generates deviations towards lower $y_{\rm eff}$.  On the other hand, a weaker SN feedback can only drive positive variations of $y_{\rm eff}$ for galaxies not significantly affected by AGN (i.e. with low $M_{\rm BH}/M_\star$); in other case, AGN feedback causes the decrease of $y_{\rm eff}$.

In this section, we discuss about the nature of our findings and their implications. We also explore our results in the context of observational data.

\subsection{The nature of deviations from the \NoAGNfp}
\label{sec:nature_deviations}

\begin{figure}
    \includegraphics[width=\columnwidth]{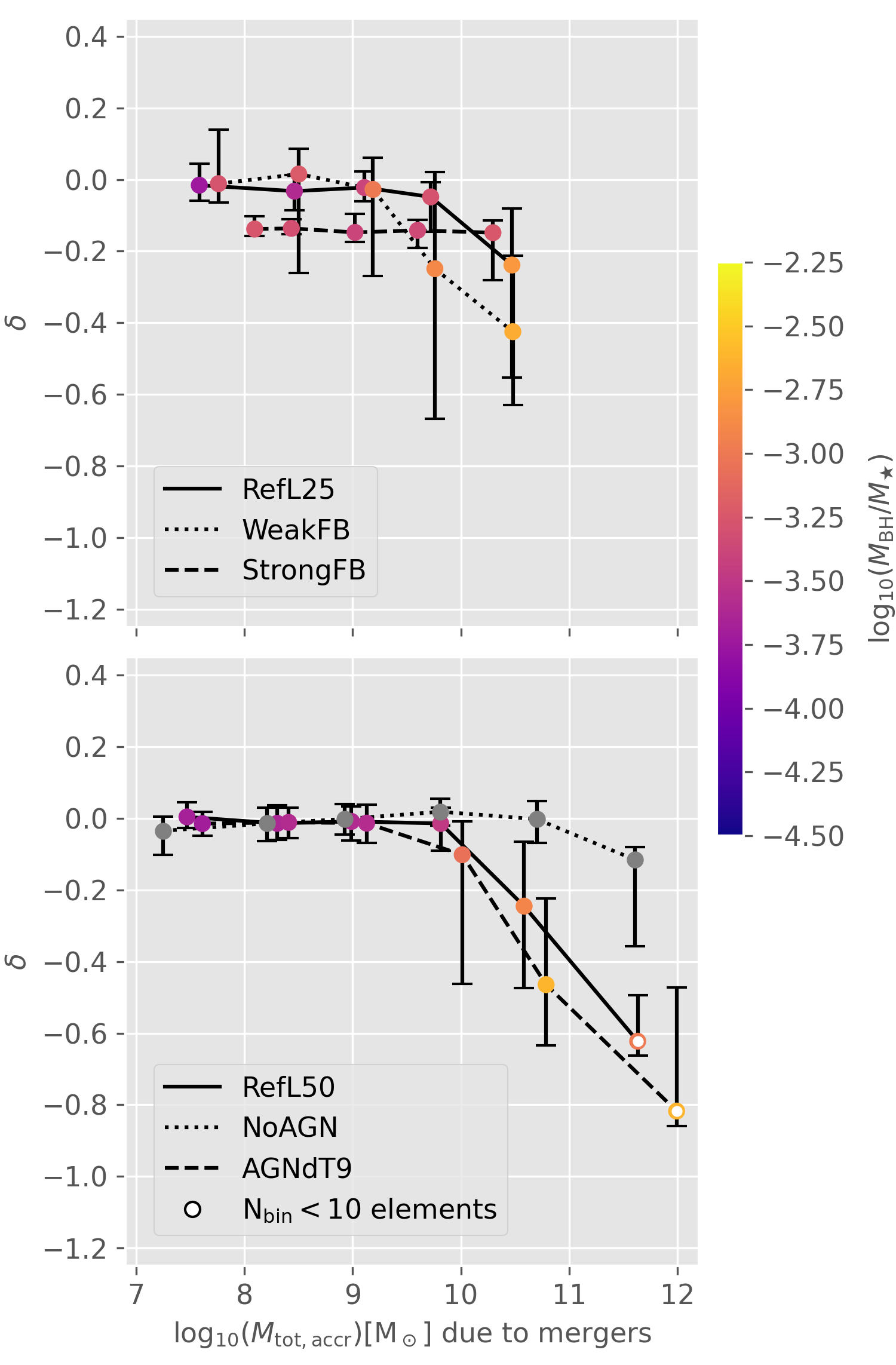}
    \caption{
    Residuals ($\delta$) from the \NoAGNfp \, as a function of the total mass (stars + SF gas + NSF gas) accreted via merger events, colour-coded by the dominance of BH, for models that varies the efficiency of SN feedback (top panel) and AGN feedback (bottom panel) (see Table~\ref{tab:simus}, for details). The curves represent the median relations, and the error bars denote the 25th and 75th percentiles. Symbols corresponding to the \NoAGN \, simulation are coloured grey, since the variable $M_{\rm BH}/M_\star$ is null for all galaxies in this model.
    }
    \label{fig:residue_accrmass}
\end{figure}

\begin{figure}
    \includegraphics[width=1\columnwidth]{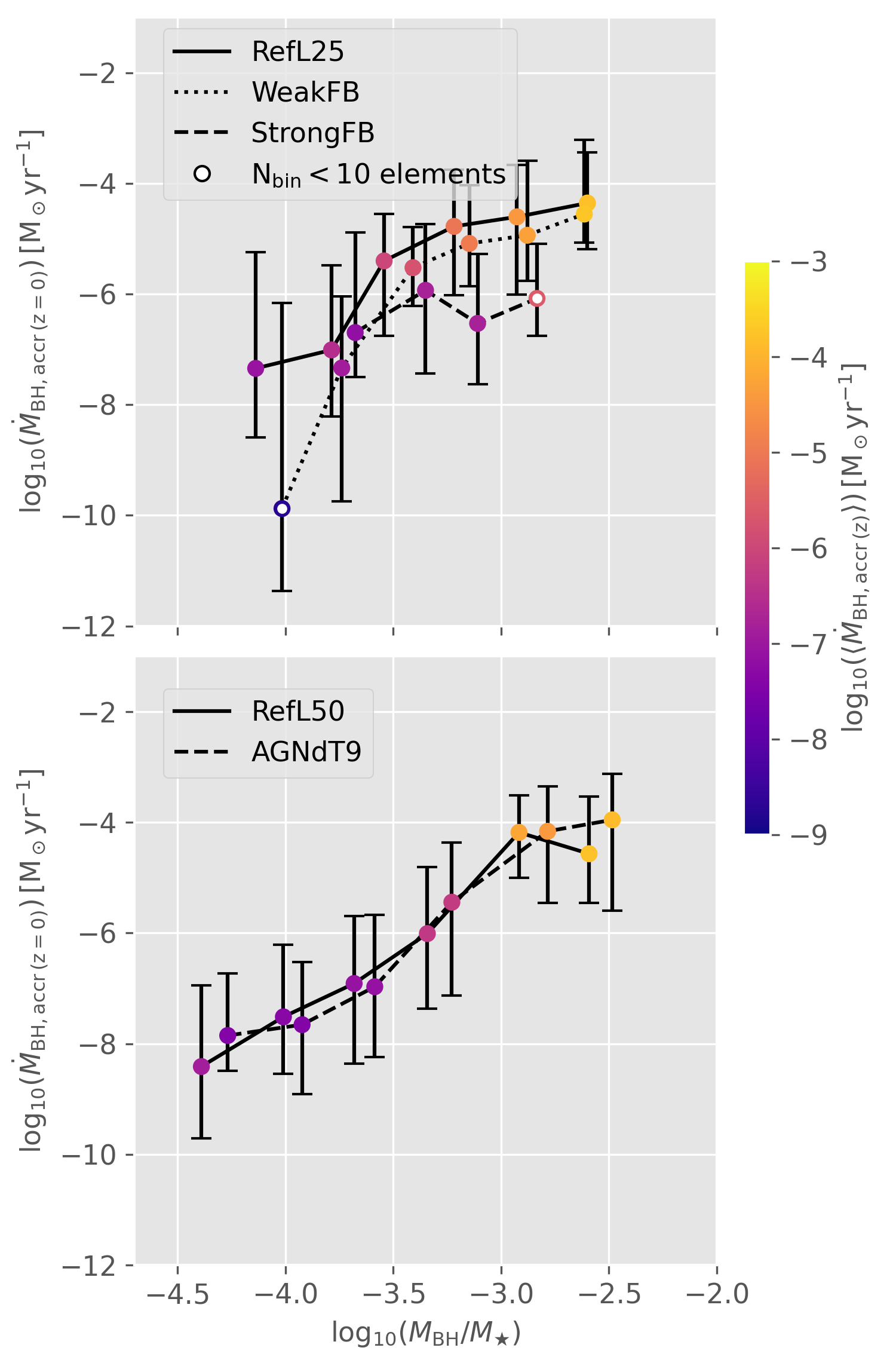}
    \caption{
    BH accretion rate at $z=0$ vs. $M_{\rm BH}/M_\star$, colour-coded by the median BH accretion rate along the galaxy lifetime calculated for the main branch of each galaxy, for models that varies the efficiency of SN feedback (top panel) and AGN feedback (bottom panel) (see Table~\ref{tab:simus}, for details). The curves represent the median relations, and the error bars denote the 25th and 75th percentiles. No data is plotted for the \NoAGN \, simulation due to the null values of $M_{\rm BH}$ for all galaxies in this model.
    }
    \label{fig:BHaccrrate_MbhMstar}
\end{figure}

In this section, we try to get more insight into the nature of the residuals  ($\delta$, see equation~\ref{eq:residuals}) from the \NoAGNfp \, (Section~\ref{sec:3d_relations}), which, in principle, seem to be strongly dependent on the dominance of the BH inside galaxies (Fig.~\ref{fig:residue_MbhMstar}). Our aim is to explore plausible important mechanisms that can enhance the growth of the BH mass in our selected galaxies and lead to such deviations.

We address the impact of merger events by analysing the formation histories of galaxies studied in the preceding sections. For this analysis, we consider mergers experienced by the main progenitor branch\footnote{The main branch of the merger tree is defined as that with the largest stellar mass summed across all earlier simulation outputs, see \cite{McAlpine2016} for more details.} of each galaxy selected at $z=0$. The results are summarised in Fig.~\ref{fig:residue_accrmass}, where $\delta$ is plotted against the total baryonic mass (including the stellar, SF gas and NSF gas components) accumulated through merger events  (i.e. $M_{\rm tot,accr} = {\sum}_i M_{{\rm bar,tot},i}$, where $i$ considers all satellites that merge onto the main progenitor along the galaxy evolution). The upper panel corresponds to simulations which adopt different SN feedback efficiencies, while the lower panel examines the impact of implementing different temperature increases due to AGN feedback. 

With the exception of the \StrongFB \, model, a consistent trend appears across the majority of models. Notably, we observe that galaxies with larger negative $\delta$ have acquired a more significant amount of baryonic mass through merger events. In addition, a more close inspection suggests also that these galaxy mergers have driven the accretion of gas and BH growth in simulated systems, which reach $z=0$ with more than 80 per cent of the total baryonic mass and more than 40 per cent of the total gas in the form of NSF gas. These findings are in agreement with the higher $M_{\rm BH}/M_\star$ shown by galaxies with $\delta < 0$: merger events can supply significant amounts of gas and potentially generate instabilities which can lead to the migration of material towards the inner galaxy regions, hence boosting the gas accretion rate onto BH and the injection of energy via AGN feedback. Moreover, mergers between BH can take place during galaxy mergers, leading to an increase in the BH mass of the remnant systems. This behaviour is not seen in the case of the \StrongFB \, model, likely due to the lower baryonic masses reached by galaxies in this sample (see Fig.~\ref{fig:yeff_vs_Mbar}); merger events are expected to have had a more significant role on the formation of more massive galaxies.

We also analysed the BH accretion rates in our $z=0$ galaxies, $\dot{M}_{\rm BH,accr \, (z=0)}$, and the median of this quantity along the galaxy lifetime calculated along the main branch of progenitors of each galaxy, $\langle \dot{M}_{\rm BH,accr \, (z)} \rangle$.\footnote{We remind that caution should be taken when studying the simulated BH accretion rates since the time sampling may not accurately capture their high temporal variability (see Section~\ref{sec:AGN_model}).} The former provides information regarding the current state of the BH, while the latter gives us details about the historical value of the BH accretion rate of each galaxy (see Section~\ref{sec:AGN_model}, for further details). This information is plotted in Fig.~\ref{fig:BHaccrrate_MbhMstar}. It is clear that, for all models, galaxies with higher $M_{\rm BH}/M_\star$ tend to show a stronger present and past average BH activity, confirming that such systems have been more significantly affected by AGN feedback.

\subsection{Gas heating impact on effective yields}
\label{sec:gas_heating}

\begin{figure*}
    \includegraphics[width=2\columnwidth]{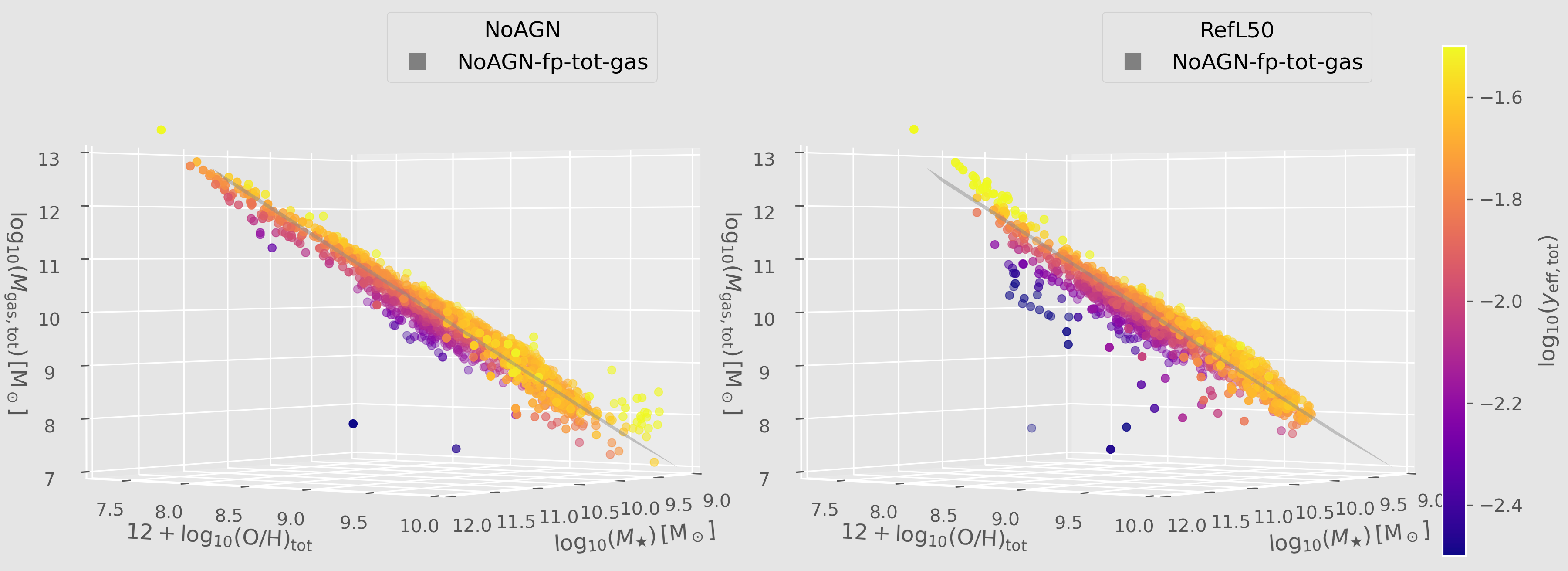}
    \caption{
    3D scatter plot of galaxies in the $M_{\star} - {\rm (O/H)_{\rm tot}} - M_{\rm gas,tot}$ space for the \NoAGN \, (left-hand panel) and \RefAGN \, (right-hand panel) simulations (see Table~\ref{tab:simus}, for details), colour-coded with $y_{\rm eff,tot}$. All quantities are calculated using the {\em total gas} component. The grey shadow represents the fitting plane obtained from the \NoAGN \, simulation. More viewing angles are available in the Supplementary Material.
    }
    \label{fig:3D_NoAGN-fp-tot-gas}
\end{figure*}

AGN and SN feedback can affect the gas component of galaxies through two main channels: 1) gas heating, which could quench the star formation and, hence, the chemical evolution of galaxies; 2) metal-enriched gas outflows, which deprive the galaxies from fuel for forming new stars and lower the metallicity of the systems (see \citealt{DeRossi2017}, for a discussion about these effects in \eagle). So far, we have defined all our gas quantities considering only the SF gas component of galaxies (Section~\ref{sec:selection}), which can vary through the two aforementioned mechanisms. But, a close inspection of our galaxy sample in different simulations reveals that galaxies below the \NoAGNfp \, tend to have significantly high percentages of NSF gas ($\gtrsim 85$ per cent, for the majority of galaxies). On the other hand, galaxies above the plane, which show the highest $y_{\rm eff}$, tend to have lower amounts of NSF gas.  These findings suggest that feedback-driven gas heating is probably the main process responsible for $y_{\rm eff}$ negative variations and deviations down the \NoAGNfp. In this context, it is interesting to explore the behaviour of galaxies if re-defining gas properties in order to include the NSF phase. Thus, we re-estimated {\em all} our galaxy properties considering the whole gas component (including SF and NSF gas) within each system. Gas heating can foster the transition from the SF to the NSF gas phase, affecting quantities derived from SF gas, while, by definition, the {\em total} gas component is not affected by such transition. Thus, the analysis of the variations of the latter component with feedback allows to probe the relevance of gas heating against that of outflows.

In Fig.~\ref{fig:3D_NoAGN-fp-tot-gas}, we analyse the 3D distribution of galaxies in the parameter space given by the total gas mass ($M_{\rm gas,tot} = M_{\rm gas,SF} + M_{\rm gas,NSF}$, where $M_{\rm gas, NSF}$ stands for the NSF gas component), $M_\star$, and oxygen abundance derived from the total gas component (${\rm O/H})_{\rm tot}$. Symbols are colour-coded with $y_{\rm eff,tot} = Z_{\rm gas,tot} / \ln(1/\mu_{\rm tot})$, where $Z_{\rm gas,tot}$ is the metallicity corresponding to the whole gas phase and $\mu_{\rm tot} = M_{\rm gas,tot} /( M_\star + M_{\rm gas,tot})$. The left and right panels show results for the \NoAGN \, and \RefAGN \, models, respectively. We see that, when taking into account the total gas component, galaxies in the \NoAGN \, simulation are again located around a plane (\NoAGNfptg, hereafter), showing even lower dispersion around the plane than that obtained in Fig.~\ref{fig:3D_plot_AGN} and \ref{fig:3D_plot_SN}. The new output parameters from the fitting of equation~\eqref{eq:plane} to the \NoAGN \, simulated sample and the corresponding scatter can be found in Table~\ref{table_fits} (second row).

The right panel of Fig.~\ref{fig:3D_NoAGN-fp-tot-gas} shows the M$_{\rm \star,g}$Z$_{\rm g}$R for the \RefAGN \, simulation. We see that, when using the whole gas component for defining our properties, negative deviations from the plane decrease significantly.  In fact, the \NoAGNfptg \, seems to constitute a good representation for almost all the galaxy population even for the \RefAGN \, model, with very few systems exhibiting very negative departures. A more in-depth examination of the latter simulation reveals that, for galaxies exhibiting the highest $M_{\rm BH}/M_\star$ ratios, the median deviation from the \NoAGNfp \, is $\delta \approx -0.5$ when calculating properties solely from the star-forming gas component. Conversely, when calculating properties from the whole gas component, the median departure from the \NoAGNfptg \, is $\delta \approx -0.25$, consistent with the smaller scatter around the plane (see Table~\ref{table_fits}). These results are in line with feedback-driven gas heating (which fosters the transition from the SF to the NSF gas phase) being the main responsible for the negative deviations from the \NoAGNfp \, (and the associated decrease of effective yields) in Fig.~\ref{fig:3D_plot_AGN} and \ref{fig:3D_plot_SN}. 

Finally, we emphasise that findings from this section do not imply that galaxies roughly behave as closed-boxes when taking into account their whole gas component.  All simulated galaxies are affected by gas inflows, outflows and mergers during their evolution. Our results point out the importance of the criteria used for characterising the gas phase for the correct use of $y_{\rm eff}$ as a tracer of feedback impact on metallicity scaling relations.

\subsection{Comparison with observations}
\label{sec:comp_obs}

\begin{figure*}
    \includegraphics[width=2\columnwidth]{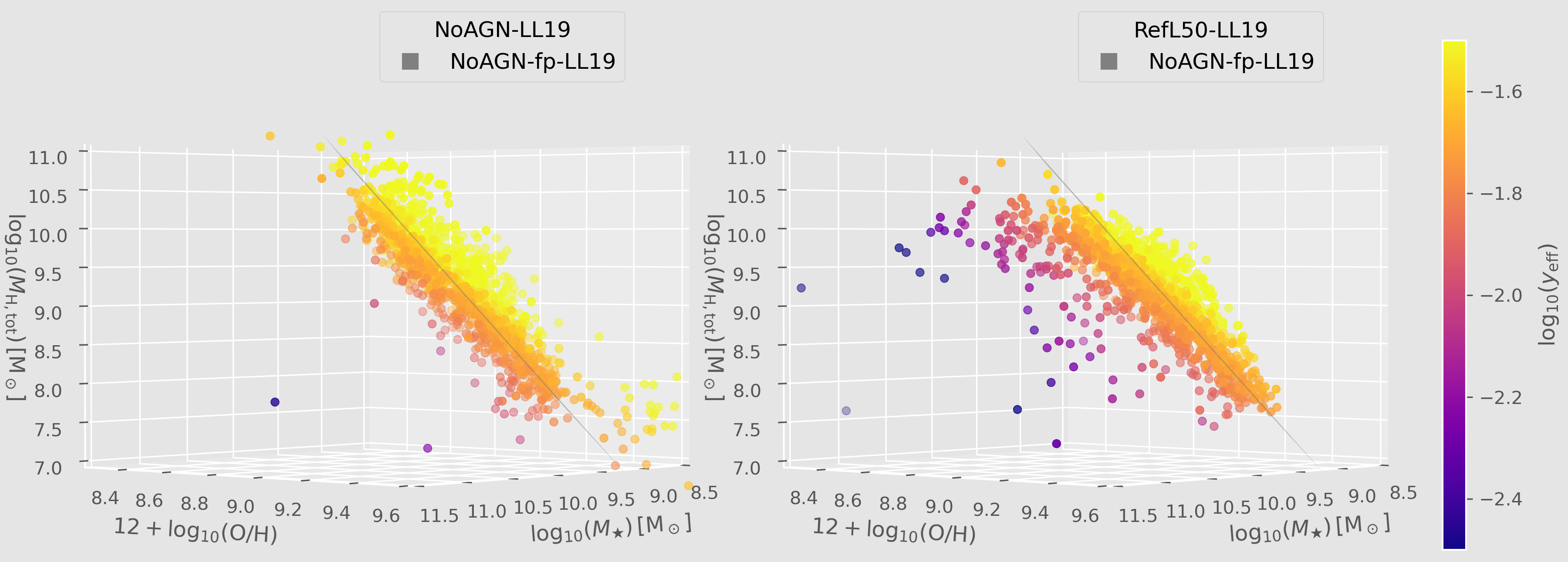}
    \caption{
    3D scatter plot in the $M_{\star} - {\rm O/H} - M_{\rm H,tot}$ space for the \NoAGN \, (left-hand panel) and \RefAGN \, (right-hand panel) simulations (see Table~\ref{tab:simus}, for details), colour-coded with $y_{\rm eff}$. All quantities are calculated using similar apertures and definitions as in \citetalias{Lara-Lopez2019}. The grey surface represents the fitting plane (edge-on view) obtained for the \NoAGN \, simulation. More viewing angles are available in the Supplementary Material.
    }
    \label{fig:3D_NoAGN-fp-LL19}
\end{figure*}

\citetalias{Lara-Lopez2019} performed a detailed comparison between effective yields in \eagle \, simulations and observations. Regarding the observational sample, they combined optical and radio wavelengths to calculate combine properties, such as gas fractions, baryonic masses and effective yields. \citetalias{Lara-Lopez2019} employed spectroscopic data from the Sloan Digital Sky Survey (SDSS), HI information from the Arecibo Legacy Fast Arecibo L-band Feed Array (ALFALFA) survey, data from the GASS and COLD GASS surveys, and a sample of star forming galaxies from the Virgo cluster. For deriving metallicities, they use the approaches in \citet{Pilyugin2016, Pilyugin2018}, applying a correction to account for oxygen locked up in dust. For more details about the estimates of observed properties, the reader is referred to \citetalias{Lara-Lopez2019}. In the case of simulations, \citetalias{Lara-Lopez2019} studied different AGN feedback models. Additionally, these authors analysed the recalibrated high-resolution run (\Recal) within the \eagle \, suite, which predicts a M$_{\star}$Z$_{\rm g}$R slope that, at low $M_\star$, agrees better with some observational works.\footnote{We acknowledge, however, that there are still many controversies regarding the exact value of the slope and zero point of the M$_{\star}$Z$_{\rm g}$R. The use of different selection criteria and methods for the estimates of stellar masses and metallicities affects the comparison between different observational data.}  However, as shown by \citetalias{Lara-Lopez2019}, the main general trends obtained for $y_{\rm eff}$ (and its associated scaling relations) are preserved when using intermediate resolution simulations run with the reference model, as those used in this article (Section~\ref{sec:simulations}).  

In order to compare \citetalias{Lara-Lopez2019} observational data with the simulation predictions presented in this article, we followed the methodology described in \citetalias{Lara-Lopez2019} to re-calculate our simulated properties in such a way that they are consistent with the instrument apertures and definitions associated with  \citetalias{Lara-Lopez2019} observations.  Very briefly, global quantities typically measured from optical diagnostics (e.g. $M_\star$, ${\rm O/H}$, $Z_{\rm gas}$) were re-estimated taking into account the mass enclosed by a spherical aperture of radius 30 kpc. As discussed in \citet{Schaye2015}, stellar masses obtained in this way are comparable to those derived from a projected circular aperture of the Petrosian radius. Besides, given that gas metallicities of the observed galaxy sample are inferred from HII SF regions, we evaluate chemical abundances by studying the SF gas component. Regarding gas mass calculations, we apply a larger aperture of $70\,{\mathrm{kpc}}$ \citep{Crain2017}, which roughly corresponds to the Arecibo L-Band Feed Array (ALFA) FWHM beam size of $\sim 3.5$ arcmin (\citealt{Giovanelli2005}) at the median redshift of the GASS sample, $z = 0.037$ (\citealt{Catinella2010}). Besides, as in \citetalias{Lara-Lopez2019}, we quantify the simulated gas mass by using the {\em total} hydrogen mass enclosed by $70\,{\mathrm{kpc}}$ ($M_{\rm H, tot}$, hereafter). 

Fig.~\ref{fig:3D_NoAGN-fp-LL19} shows the M$_{\rm \star,g}$Z$_{\rm g}$R corresponding to simulations \NoAGN \, (left-hand panel) and \RefAGN \, (right-hand panel), obtained when using quantities calculated consistently with the \citetalias{Lara-Lopez2019} observations, as explained above.  Again, galaxies corresponding to the \NoAGN \, model tend to be located around a plane, which seems to roughly describe a surface with constant $y_{\rm eff}$.  In Table~\ref{table_fits} (third row), we report the output parameters from the fitting of equation~\eqref{eq:plane} to the latter plane (\NoAGNfp-LL19, hereafter). Once more, we obtain that larger negative (positive) deviations from the plane can be associated with lower (higher) $y_{\rm eff}$. The bulk of galaxies in the \RefAGN \, simulation also aligns closely with the latter plane, with larger deviations observed in galaxies with the lowest $y_{\rm eff}$. Considering the scatter around the \NoAGNfp-LL19\, ($ \epsilon \approx 0.3$, Table~\ref{table_fits}), we can predict a characteristic value ${\delta}_{\rm 0,obs} = -0.3$ ($y_{\rm eff} \approx 10^{-2.4}$), such that observed galaxies with residuals ${\delta} \la {\delta}_{\rm 0,obs}$ ($y_{\rm eff} \la 10^{-2.4}$) have been probably affected by a strong AGN feedback.

\begin{figure}
    \includegraphics[width=\columnwidth]{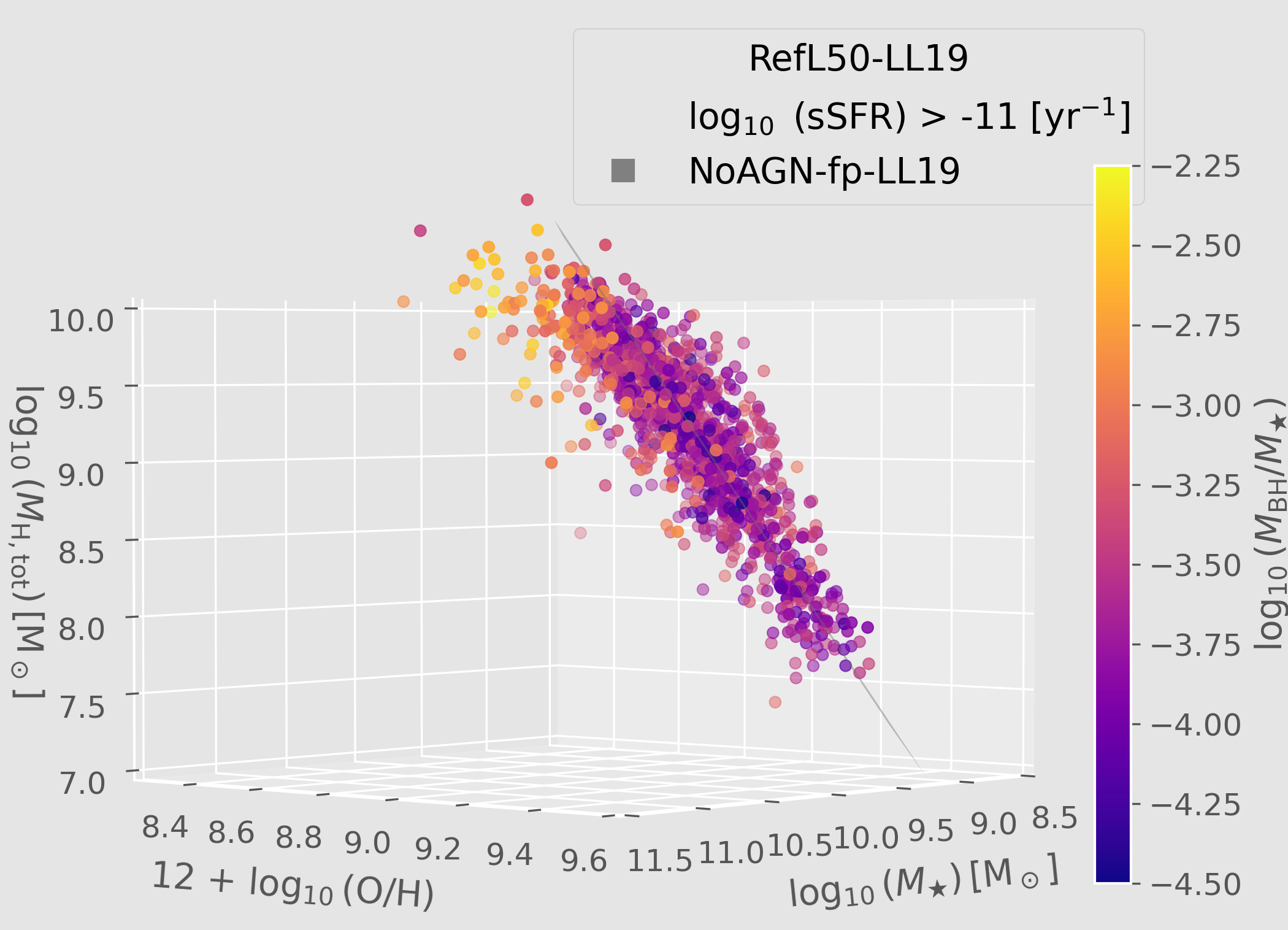}
    \caption{
    3D scatter plot in the $M_{\star} - {\rm O/H} - M_{\rm gas}$ space for the \RefAGN \, simulation. Only galaxies with ${\rm sSFR} > 10^{-11} [{\rm yr}^{-1}]$ are plotted. The colour bar indicates the $M_{\rm BH}/M_\star$ of each galaxy. All quantities are calculated using similar apertures and definitions as in \citetalias{Lara-Lopez2019}. The grey surface represents the best fitting plane (edge-on view) obtained for the \NoAGN \, simulation. To explore other viewing angles, check the Supplementary Material.    
    }
    \label{fig:sSFR_threshold}
\end{figure}

For the sake of comparison with observational samples constituted by SF galaxies, we re-analysed the M$_{\rm \star,g}$Z$_{\rm g}$R corresponding to the \RefAGN \, simulation (Fig.~\ref{fig:3D_NoAGN-fp-LL19}, right panel) for systems with ${\rm sSFR} > 10^{-11} [{\rm yr}^{-1}]$. Results are shown in Fig.~\ref{fig:sSFR_threshold}. It is clear that very large deviations from the \NoAGNfp-LL19 (edge-on grey surface) are not expected for star-forming galaxies. These trends align with the notion that such deviations are primarily generated by galaxies that are more affected by AGN feedback, as extensively discussed in previous sections. Nevertheless, we would like to notice that, although host galaxies currently showing AGN activity are usually removed when studying metallicity scaling relations, AGN feedback might have occurred in cyclic episodes that affected the progenitors of observed galaxy populations. In \eagle, AGN activity also evolves with time and all BH are actively accreting gas some part of the time, giving place to the AGN phenomenon.
In other words: although the accretion rate onto the BH could be low today, the BH activity could have been more significant in the past.  This can be seen in Fig.~\ref{fig:BHaccrrate_MbhMstar}, where it is clear that the BH accretion rates at $z=0$ ($\dot{M}_{\rm BH,accr \, (z=0)}$) can be much lower than the median of this quantity along the galaxy lifetime ($\langle \dot{M}_{\rm BH,accr \, (z)} \rangle$).  In the particular case of galaxies shown in Fig.~\ref{fig:sSFR_threshold}, we find that, for $\approx 50 \%$ of the sample, $\dot{M}_{\rm BH,accr \, (z=0)} / \langle \dot{M}_{\rm BH,accr \, (z)} \rangle < 1$.

\begin{figure*}
    \includegraphics[width=2\columnwidth]{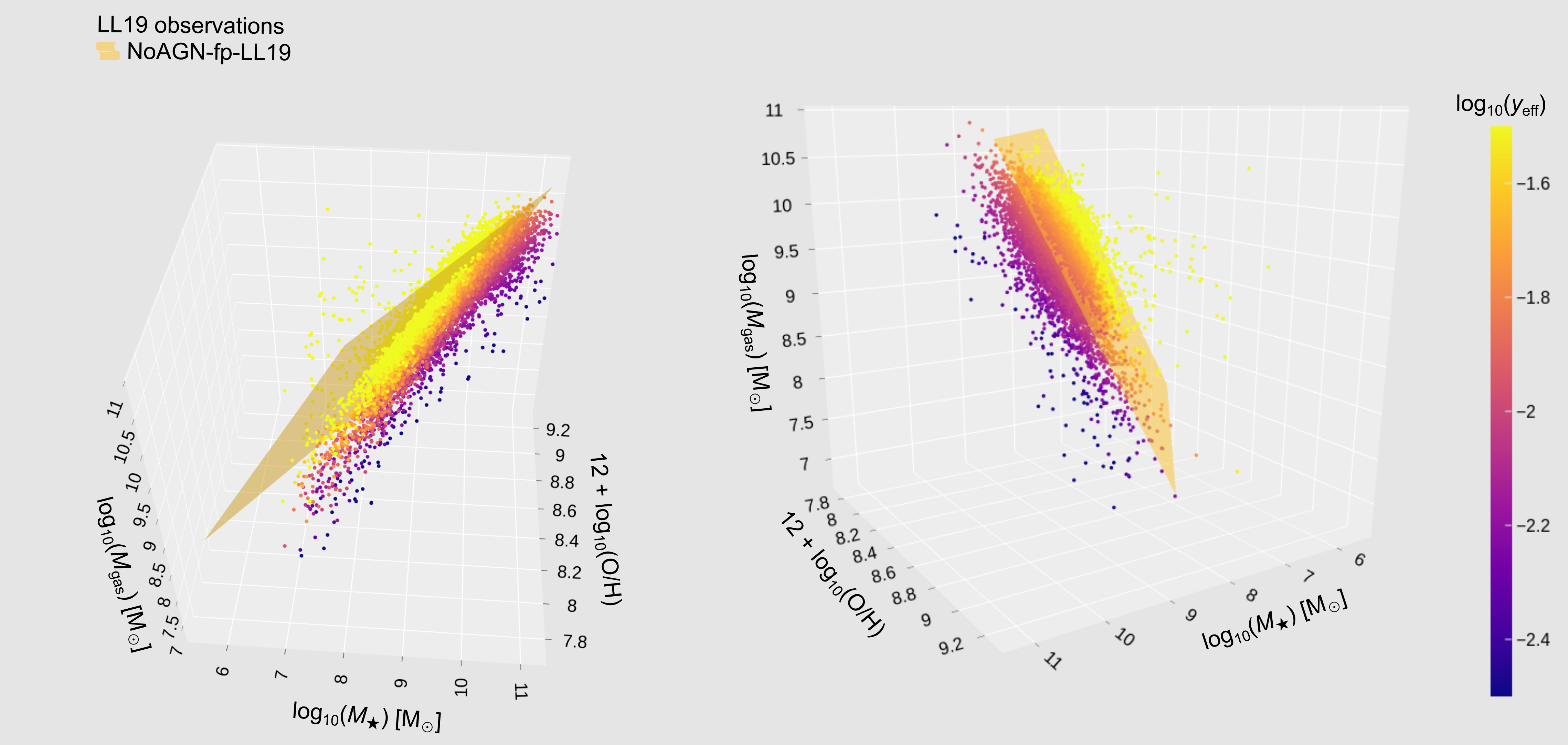}
    \caption{
    3D scatter plot in the $M_{\star} - {\rm O/H} - M_{\rm gas}$ space for the \citetalias{Lara-Lopez2019} observational data, colour-coded with $y_{\rm eff}$. The yellow surface represents the best fitting plane obtained for the \NoAGN \, simulation, when quantities are calculated using similar apertures and definitions as in \citetalias{Lara-Lopez2019} (see text for details). We encourage the reader to see the Supplementary Material for more viewing angles.
    }
    \label{fig:3D_LL19_NoAGN-fp-LL19}
\end{figure*}

Finally, in Fig.~\ref{fig:3D_LL19_NoAGN-fp-LL19}, we analyse the \citetalias{Lara-Lopez2019} {\em observational} sample in the 3D parameter space given by gas mass, stellar mass and ${\rm O/H}$.\footnote{The reader is refereed to \citetalias{Lara-Lopez2019} for a description of the procedures used for deriving galaxy quantities in observations.} Encouragingly, the observed galaxy sample is located around our previously defined NoAGN-fp-LL19, which is shown as a yellow surface in the figure. The good agreement between observations and \eagle \, simulations regarding the behaviour of effective yields has already been noted by \citetalias{Lara-Lopez2019}. The highest (lowest) observed $y_{\rm eff}$ can be associated to galaxies above (below) the plane, with the residuals $\delta$ from such a plane, tracing variations in $y_{\rm eff}$.  And, according to findings in this paper, the behaviour of $y_{\rm eff}$ (and, hence, of $\delta$) is significantly modulated by the joint accumulated effects of SN and AGN feedback. Although disentangling the impact of the latter processes could not be plausible in many cases, the location of galaxies with respect to the plane can provide some hints about their feedback histories, as we discuss in the next section.

\subsection{Connecting the features of the M$_{\rm \star,g}$Z$_{\rm g}$R to the plausible feedback histories of galaxies}

Our findings suggest a close connection between feedback, $y_{\rm eff}$ and the location of galaxies in the 3D $M_{\rm \star}$-O/H-$M_{\rm gas}$ parameter space, which could be useful for proposing feedback scenarios for real galaxy populations. 

As extensively discussed, simulated galaxies that were not significantly affected by AGN feedback ($M_{\rm BH}/M_{\star}  \approx 0$), are well represented by a plane in the aforementioned 3D space, with this plane located along a region of an almost constant high $y_{\rm eff}$.  
Interestingly, the existence of such a plane seems to be robust against the different variations that we implemented for defining our main properties (i.e. gas mass, stellar mass and O/H), as explained in previous sections.  But, our different definitions lead to moderate changes in the normalisation and orientation of the plane (Table~\ref{table_fits}). In particular, when defining the properties of simulated galaxies according to those of SF galaxies in the Local Universe, galaxies in the latter sample are located around the plane derived from simulations.

Our findings so far suggest that the location of galaxies around such characteristic plane could provide clues about the feedback histories of real galaxies, as indicated below:

\begin{itemize}
    \item Galaxies that are well represented by the plane show an almost constant high $y_{\rm eff}$ and nearly null residuals $\delta$.  Such galaxies are not expected to have been significantly affected by AGN feedback ($M_{\rm BH}/M_{\star}  \approx 0$).
    \item Considering the scatter obtained for the plane \, (e.g. $\epsilon$ in Table~\ref{table_fits}), galaxies with residuals $\delta$ below such characteristic value can be considered more affected by the accumulated effects of AGN feedback. Such galaxies are expected to display very high $M_{\rm BH}/M_{\star}$ and very low $y_{\rm eff}$, compared with the median population located closer to the plane. 
    \item For a reference AGN feedback efficiency, galaxies more strongly affected by SN feedback tend to lie slightly below the \NoAGNfp \, ($\delta \lesssim 0$) and does not reach very high $y_{\rm eff}$ values. Thus, galaxies with the highest $y_{\rm eff}$ (those above the plane) cannot be regarded as systems strongly affected by SN feedback. 
    \item For a reference AGN feedback efficiency but a weak SN feedback, we detect two different trends: 1) galaxies with  higher $M_{\rm BH}/M_{\star}$ show lower $y_{\rm eff}$ and negative distances to the plane ($\delta < 0$); 2) galaxies with lower $M_{\rm BH}/M_{\star}$ have higher $y_{\rm eff}$ and tend to be located above the \NoAGNfp \, ($\delta \gtrsim 0$). Hence, galaxies with the highest $y_{\rm eff}$ (those above the plane) can be associated to systems with both, a weak SN feedback efficiency and a weak AGN feedback impact. 
\end{itemize}

We note that we do not have information regarding black hole masses for our observational sample. Future determination of these masses would be very useful to test the predictions of our simulations. 

Finally, it is worth acknowledging that the comparison between simulations and observations could not be so straightforward.
As discussed in \cite{DeRossi2017}, the normalisation of the observed M$_{\star}$Z$_{\rm g}$R is still a matter of discussion, with different observational works reporting a variety of results due to the use of distinct metallicity calibrators.  In addition, uncertainties in the nucleosynthetic yields implemented in simulations can also affect the determination of absolute metallicity values. This issue regarding the calibration of absolute metallicities affects also the estimate of $y_{\rm eff}$ (equation~\ref{eq:yeff}). In this context, the comparison of global metallicities and $y_{\rm eff}$ between simulations and observations should be addressed with care.

\section{Summary and Conclusions}
\label{sec:summary}
We have analysed the effective yields ($y_{\rm eff}$) of galaxies in \eagle \, cosmological hydrodynamical simulations that implement different SN and AGN feedback prescriptions: reference models calibrated against some observations (simulations \RefAGN \, and \RefSN); a model with weaker SN feedback (\WeakFB \, simulation), and another with stronger SN feedback (\StrongFB \, simulation); a simulation without BH and, hence, without AGN feedback (\NoAGN \, simulation); and a simulation which applies a higher gas temperature increment ($\Delta T_{\rm AGN}$) associated to AGN feedback (\AGNdT \, simulation), which drives a stronger AGN feedback impact according to previous works. In particular, we try to evaluate the role of SN and AGN accumulated feedback effects on the determination of $y_{\rm eff}$ and the features of its associated scaling relations. For our main analysis, galaxy properties whose definitions involve gas (e.g. chemical abundances, gas fractions and $y_{\rm eff}$) were derived from the SF gas-phase. But, we also explored the effects of using other plausible definitions. Our most relevant findings and conclusions can be summarised as follows:

\begin{itemize}
    \item  In agreement with previous works, the reference model predicts an average positive (negative) slope for the $y_{\rm eff} - M_{\rm bar}$ relation (Fig.~\ref{fig:yeff_vs_Mbar}) at low $M_{\rm bar} \la 10^{10}~{\rm M}_{\odot}$ (high $M_{\rm bar} \ga 10^{10}~{\rm M}_{\odot}$). An increase of $\Delta T_{\rm AGN}$ generates a decrease of such a slope at high masses but has a no significant impact for low-mass systems. Interestingly, similar effects are obtained when applying a weaker SN feedback efficiency, given that it fosters the formation of a significant number of galaxies with dominant BH (i.e., with high BH mass-stellar mass ratio, $M_{\rm BH} / M_{\star}$). On the other hand, no dominant BH are formed in the case of a strong SN feedback scenario. In comparison with the reference model, a stronger SN feedback leads to an overall decrease of $y_{\rm eff}$ at all masses and, also, predicts a positive slope for the $y_{\rm eff} - M_{\rm bar}$ for the whole simulation mass range. 
    \item To understand the origin of $y_{\rm eff}$ behaviour, we analysed the feedback impact on the key galaxy properties involved in the $y_{\rm eff}$ definition: gas-phase metallicity, stellar mass and gas mass (Fig.~\ref{fig:2D_AGN_models} and \ref{fig:2D_SN_models}). AGN feedback seems to drive the simultaneous decrease of the SF gas fraction and metallicity of massive galaxies (which have more dominant BH), which explains their lower $y_{\rm eff}$. On the other hand, a stronger SN feedback efficiency favours the decrease of metallicity at fixed SF gas fraction, also leading to lower $y_{\rm eff}$. In the case of a weak SN feedback efficiency, a more complex behaviour arises because of the formation of BH in a significant number of galaxies along our whole mass range:  as a consequence of AGN feedback, galaxies with dominant BH show lower ${\rm O/H}$, lower SF gas fractions and, hence, lower $y_{\rm eff}$, while galaxies with low $M_{\rm BH}/M_{\star}$ tend to have higher metallicities at a given SF gas fraction and, thus, higher $y_{\rm eff}$.
    \item We found a clear anti-correlation between $y_{\rm eff}$ and the $M_{\rm BH} / M_\star$ for galaxies in all studied feedback models (Fig.~\ref{fig:yeff_vs_MbhMstar}): galaxies with $M_{\rm BH} / M_\star \la 10^{-3.5}$ show a higher and roughly constant $y_{\rm eff}$ while, for galaxies with $M_{\rm BH} / M_\star  \ga 10^{-3.5}$, $y_{\rm eff}$ strongly decreases with $M_{\rm BH} / M_\star$.     
    \item We also studied the distribution of galaxies in the 3D parameter space defined by gas-phase metallicity, $M_\star$ and gas mass (M$_{\rm \star,g}$Z$_{\rm g}$R) for different SN and AGN feedback models (Fig.~\ref{fig:3D_plot_AGN} \-- \ref{fig:3D_plot_SN}). Interestingly, we found that, when AGN feedback is turned off, the M$_{\rm \star,g}$Z$_{\rm g}$R is well described by a plane (\NoAGNfp), which is locally situated along a surface of constant $y_{\rm eff}$.  Such a plane can roughly describe the behaviour of galaxies with no dominant BH (low $M_{\rm BH} / M_{\star}$) in all simulations, regardless of the implemented feedback model. Orthogonal departures from the plane ($\delta$) can be directly associated to variations in $y_{\rm eff}$ (i.e. different $y_{\rm eff}$ isosurfaces are crossed when moving away from the plane). Deviations from the plane that generates an increase of $y_{\rm eff}$ are regarded as `positive' (i.e. galaxies are considered to be above the plane) in this work, and `negative' (i.e. galaxies are considered to be below the plane), otherwise.
    \item Galaxies highly affected by the accumulated effects of AGN feedback (i.e. systems with high $M_{\rm BH} / M_{\star}$) show very low $y_{\rm eff}$ and present large negative deviations from the \NoAGNfp \, (Fig.~\ref{fig:yeff_vs_MbhMstar}, \ref{fig:residue_MbhMstar}). If galaxies are not significantly affected by AGN feedback ($M_{\rm BH}/M_{\star}  \approx 0$), they are located close to the \NoAGNfp, having an almost constant high value of $y_{\rm eff}$. 
    \item For a reference AGN feedback efficiency, a strong SN feedback generates a moderate displacement of the bulk galaxy population downward the \NoAGNfp, with such galaxies showing intermediate $y_{\rm eff}$ values. For a reference AGN feedback efficiency but a weak SN feedback, we detect two different regimes: 
    1) galaxies with  higher $M_{\rm BH}/M_{\star}$ show lower $y_{\rm eff}$ and negative deviations from the plane; 
    2) galaxies with lower $M_{\rm BH}/M_{\star}$ have higher $y_{\rm eff}$ and tend to be located above the \NoAGNfp \, (Fig.~\ref{fig:yeff_vs_MbhMstar}, \ref{fig:residue_MbhMstar}). 
    \item A deeper analysis suggests that systems with larger negative $\delta$ present higher $M_{\rm BH} / M_{\star}$ as a consequence of their more significant amount of mass accreted via galaxy mergers (Fig.~\ref {fig:residue_accrmass}).  This is because galaxy mergers tend to boost the gas accretion rate of the BH and, also, can drive BH mergers. The present and past average BH accretion rates of galaxies with higher $M_{\rm BH} / M_{\star}$ are also higher, indicating that such systems have been more significantly affected by AGN feedback during their evolution (Fig.~\ref{fig:BHaccrrate_MbhMstar}).
    \item
    Feedback-driven gas heating seems to be the main process responsible for larger $y_{\rm eff}$ negative variations and higher deviations down the \NoAGNfp. Gas heating foster the transition from the SF to the non-SF (NSF) gas phase, affecting all our quantities derived from SF gas. If re-estimating these galaxy properties by using the {\em total} gas component (including SF and NSF gas), almost the whole galaxy population in all studied simulations is located around a new plane in the 3D space given by gas-phase metallicity, $M_\star$ and gas mass (Fig.~\ref{fig:3D_NoAGN-fp-tot-gas}). The scatter around this plane is lower and no significant negative deviations from it are obtained.
    \item The M$_{\rm \star,g}$Z$_{\rm g}$R of observed galaxies in the Local Universe also seems to be located around a plane, with the highest (lowest) effective yields associated to galaxies above (below) the plane. If re-calculating our simulated quantities mimicking the observational definitions, the observed and simulated trends show very good agreement. In particular, the new plane derived from the \NoAGN \, model (\NoAGNfp-LL19, Fig.~\ref{fig:3D_NoAGN-fp-LL19}) by using such quantities represents again a surface of constant $y_{\rm eff}$ and, encouragingly, this plane can characterise well the behaviour of observed galaxies with similar $y_{\rm eff}$ (Fig.~\ref{fig:3D_LL19_NoAGN-fp-LL19}). In addition, if restricting our simulated galaxy sample to systems with ${\rm sSFR} > 10^{-11} [{\rm yr}^{-1}]$ as in observations of star-forming galaxies, large negative $\delta$ are not obtained (Fig.~\ref{fig:sSFR_threshold}).   
\end{itemize}

To sum up, according to the findings in this paper, the evolution of $y_{\rm eff}$ (and, hence, of $\delta$) is significantly affected by the joint accumulated effects of SN and AGN feedback. Thus, the determination of $y_{\rm eff}$ (or, equivalently, $\delta$) and its associated scaling relations could provide relevant hints regarding the different feedback histories of real galaxy populations. Nevertheless, as discussed, caution should be taken when estimating the quantities involved in the definition of $y_{\rm eff}$, for a correct interpretation of observational data. In a forthcoming article, we will extend the present study, addressing the effects of feedback on $y_{\rm eff}$ at $z>0$ (for preliminary results, see \citealt{Zerbo2022}).

\section*{Acknowledgements}
We thank the reviewer for the constructive suggestions and comments that helped improve this paper. MCZ thanks {\it Asociaci\'on Argentina de Astronom\'{\i}a} for having been awarded with a grant, which partially supported this project. We thank Rodrigo Flores-Freitas for his useful help with some technical details  regarding the preparation of the Supplementary Material. SAC acknowledges funding from CONICET (PIP-2876), {\it Agencia Nacional de Promoci\'on de la Investigaci\'on, el Desarrollo Tecnol\'ogico y la Innovaci\'on} (Agencia I+D+i, PICT-2018-3743), and the {\it Universidad Nacional de La Plata} (G11-150), Argentina. We acknowledge support from PICT-2021-GRF-TI-00290 of Agencia I+D+i (Argentina). We acknowledge the Virgo Consortium for making their simulation data available. The EAGLE simulations were performed using the DiRAC-2 facility at Durham, managed by the ICC, and the PRACE facility Curie based in France at TGCC, CEA, Bruy\`{e}res-le-Ch\^{a}tel. This work used the DiRAC@Durham facility managed by the Institute for Computational Cosmology on behalf of the STFC DiRAC HPC Facility (www.dirac.ac.uk). The equipment was funded by BEIS capital funding via STFC capital grants ST/P002293/1, ST/R002371/1 and ST/S002502/1, Durham University and STFC operations grant ST/R000832/1. DiRAC is part of the National e-Infrastructure.

\section*{Data Availability}
The \eagle \, simulations are publicly available. Both halo/galaxy catalogues and particle data can be accessed and downloaded at \url{www.icc.dur.ac.uk/Eagle/} \citep{Schaye2015,Crain2015,McAlpine2016}. To create the results shown in this publication, Python libraries were used (\textsc{AstroPy}, \textsc{NumPy}, \textsc{SciPy}, \textsc{PyPlot}, and \textsc{H5Py}), including the publicly available \textsc{read\_eagle} module (\url{https://github.com/jchelly/read_eagle}, \citealp{eagle2017}). Additional data and code directly related to this work are available on reasonable request from the corresponding author.

\bibliographystyle{mnras}
\bibliography{Zerbo-bib} 







\bsp	
\label{lastpage}
\end{document}